\documentclass[%
reprint,
nofootinbib,
 amsmath,amssymb,
 aps,
]{revtex4-2}

\usepackage{graphicx}% Include figure files
\usepackage{dcolumn}% Align table columns on decimal point
\usepackage{bm}% bold math
\usepackage{physics}
\usepackage{amsmath}
\usepackage{color}
\usepackage{siunitx}
\usepackage{placeins}

\newcommand{\x}{\vb{z}}

\newcommand{\cD}{{\mathcal D}}

\newcommand{\bs}[1]{\boldsymbol{#1}}

\newcommand{\DD}{\bs{\mathcal{D}}}

\begin{document}

\preprint{APS/123-QED}
\title{Detection of diffusion anisotropy from an individual short particle trajectory}% Force line breaks with \\

\author{Kaito Takanami}
\email{takanami255@g.ecc.u-tokyo.ac.jp}
\affiliation{Department of Physics, Graduate School of Science, The University of Tokyo, Tokyo 113-0033, Japan}

\author{Daisuke Taniguchi}
\affiliation{International Research Center for Neurointelligence (WPI-IRCN), The University of Tokyo, Tokyo 113-0033, Japan}
\affiliation{Laboratory for Cell Polarity Regulation, RIKEN Center for Biosystems Dynamics Research (BDR), Osaka 565-0874, Japan}

\author{Sawako Enoki}
\affiliation{Universal Biology Institute (UBI), The University of Tokyo, Tokyo 113-0033, Japan}
\affiliation{Laboratory for Cell Polarity Regulation, RIKEN Center for Biosystems Dynamics Research (BDR), Osaka 565-0874, Japan}

\author{Masafumi Kuroda}
\affiliation{International Research Center for Neurointelligence (WPI-IRCN), The University of Tokyo, Tokyo 113-0033, Japan}
\affiliation{Laboratory for Cell Polarity Regulation, RIKEN Center for Biosystems Dynamics Research (BDR), Osaka 565-0874, Japan}

\author{Yasushi Okada}
\affiliation{Department of Physics, Graduate School of Science, The University of Tokyo, Tokyo 113-0033, Japan}
\affiliation{International Research Center for Neurointelligence (WPI-IRCN), The University of Tokyo, Tokyo 113-0033, Japan}
\affiliation{Department of Cell Biology, Graduate School of Medicine, The University of Tokyo, Tokyo 113-0033, Japan}
\affiliation{Universal Biology Institute (UBI), The University of Tokyo, Tokyo 113-0033, Japan}
\affiliation{Laboratory for Cell Polarity Regulation, RIKEN Center for Biosystems Dynamics Research (BDR), Osaka 565-0874, Japan}

\author{Yoshiyuki Kabashima}
\email{kaba@phys.s.u-tokyo.ac.jp}
\affiliation{Department of Physics, Graduate School of Science, The University of Tokyo, Tokyo 113-0033, Japan}
\affiliation{The Institute for Physics of Intelligence, The University of Tokyo, Tokyo 113-0033, Japan}
\affiliation{Trans-Scale Quantum Science Institute, The University of Tokyo, Tokyo 113-0033, Japan}

\date{\today}% It is always \today, today,

\begin{abstract}
In parallel with advances in microscale imaging techniques, the fields of biology and materials science have focused on precisely extracting particle properties based on their diffusion behavior. Although the majority of real-world particles exhibit anisotropy, their behavior has been studied less than that of isotropic particles. In this study, we introduce a new method for estimating the diffusion coefficients of individual anisotropic particles using short-trajectory data on the basis of a maximum likelihood framework. Traditional estimation techniques often use mean-squared displacement (MSD) values or other statistical measures that inherently remove angular information. Instead, we treated the angle as a latent variable and used belief propagation to estimate it while maximizing the likelihood using the expectation-maximization algorithm. Compared to conventional methods, this approach facilitates better estimation of shorter trajectories and faster rotations, as confirmed by numerical simulations and experimental data involving bacteria and quantum rods. Additionally, we performed an analytical investigation of the limits of detectability of anisotropy and provided guidelines for the experimental design. In addition to serving as a powerful tool for analyzing complex systems, the proposed method will pave the way for applying maximum likelihood methods to more complex diffusion phenomena.

\end{abstract}

\maketitle

%\tableofcontents

\section{Introduction}
The advances in imaging techniques have made it possible to visualize the spatiotemporal dynamics of various materials in microscales. This has sparked the development of theoretical methods for quantitatively extracting the physical properties of materials from experimental data. Among these methods, analyzing the mobility of a single tracer particle immersed in viscoelastic material has emerged as a powerful technique to reveal the microenvironment of the materials~\cite{Windows-Yule2022-lk}. To date, this technique has been widely applied to elucidate the dynamics and structure of cell membranes~\cite{Saxton1997-ar, Manzo2015-tx, Wang2021-vu}, DNA synthesis ~\cite{Liao2016-sm, Ma2022-yd}, principles of protein transport ~\cite{Yang2019-dw, Holcman2018-th, Cui2018-mj}, and even the processes and infection mechanisms of viruses~\cite{Liu2020-nh, Liu2019-uf}.

Conventionally, this technique often assumes the use of spherical tracer particles ~\cite{Bullerjahn2021-zz} or nonrotational tracer particles ~\cite{Lin2021-ne} for ease of statistical analysis. In other words, 
the difficulty of statistical analysis has prevented the application of this method to complex materials such as biological samples, where tracer particles generally have non-spherical shapes. However, recent biological studies have revealed the numerous characteristics and roles of anisotropic particles~\cite{Pradhan2022-ue, Cai2021-up, Pearce2021-gd, Mirza2020-dq, Meyer2016-gv}. For example, their unique shapes and properties have been shown to be useful in drug delivery~\cite{Lee2012-hc, Yang2019-qi, Kargari_Aghmiouni2023-gc} and in the control of critical phenomena in active matter~\cite{Moran2022-ya}.
This had led to a growing interest in the potential and versatility of anisotropic particles in biological research.

Despite the growing interest in these particles, to date, only a few theoretical methods have been proposed to address the trajectories of non-spherical tracer particles~\cite{han2006brownian,ribrault2007diffusion,roh2015diffaniso,hanasaki2012detection,Yoshida2006-fn}. These methods often require an unrealistically large number of timepoints in a single trajectory or the averaging of a large number of multiple trajectories. Moreover, some studies have assumed that the orientation of anisotropic particles can be observed~\cite{Charsooghi2011-zs}. However, this assumption is often invalid, especially when the particles are very small.

Taking these circumstances into consideration, this study addresses the practical problem of estimating the diffusion coefficients of a two-dimensional anisotropic particle from single-particle tracking (SPT) data. To this end, we employed the stochastic model of anisotropic diffusion introduced in~\cite{han2006brownian}. Consider a two-dimensional particle with different translational diffusion coefficients $D_a > D_b$, which correspond to the major and minor axis directions, respectively. In addition, let $D_\theta$ be the rotational diffusion coefficient of the particle. The dynamics of the center-of-mass position coordinates $x,y$ and angles $\theta$ of a single particle are described by the following Langevin equation:
\begin{align}
  \pdv{x}{t} = \xi_1(t) ~~ \pdv{y}{t} =  \xi_2(t) ~~ \pdv{\theta}{t} = \xi_3(t) \label{Lan}
\end{align}
where $\xi_1(t), \xi_2(t), \xi_3(t)$ are Gaussian white noise characterized by $\expval{\xi_1(t)}=\expval{\xi_2(t)}=\expval{\xi_3(t)}=0$ and
    \begin{eqnarray*}
      \langle \xi_i(t) \xi_j(t') \rangle &=& \Sigma_{ij} \delta(t-t') ~~ (i,j=1,2)\\
      \langle \xi_3(t) \xi_3(t') \rangle &=& 2D_\theta \delta(t-t') 
    \end{eqnarray*}
where $\vb{\Sigma}$ denotes  
    \begin{eqnarray*}
      \vb{\Sigma}(\theta)= \mqty(2 \bar{D}+\Delta D \cos{2 \theta} & \Delta D \sin{2\theta} \\ \Delta D\sin{2\theta} & 2 \bar{D}-\Delta D \cos{2 \theta}) 
    \end{eqnarray*}
where $\bar{D} = (D_a + D_b)/2$ and $\Delta D = D_a - D_b$. Brackets represent the average with respect to $\xi_1(t), \xi_2(t), \xi_3(t)$. Furthermore, when Gaussian observation noise is added to each coordinate, the observed coordinates $X,Y$ are given by $X(t) = x(t) + \epsilon \xi_4(t)$ and $Y(t) = y(t) + \epsilon \xi_5(t)$, where $\epsilon$ is the standard deviation of the noise magnitude, and $\xi_4(t), \xi_5(t)$ represent independent standard Gaussian white noise. Using these notations, we aim to precisely estimate the diffusion coefficients $D_a$, $D_b$, and $D_\theta$ from a pair of observed time series, $ \vb{X}, \vb{Y} = \{X(t_i), Y(t_i)\}_{1\leq i\leq N}$ with a time interval $\Delta t$. Note that we consider a general situation in which the orientation of the particles cannot be observed and must be inferred.
Mean-squared displacement is a classical approach that is still the most commonly used technique to estimate diffusion coefficients 
(MSD)~\cite{Qian1991-sh, Weimann2013-nd, Ernst2013-qc, Saxton1997-jo, Vestergaard2014-vp}. Two well-known methods are used for applying MSD fitting to particle trajectories obtained by SPT~\cite{Saxton1997-jo}: one is a trajectory-segmentation method that allows overlap, and the other is a method that extracts trajectories independently without allowing overlap. However, in two-dimensional anisotropic diffusion, anisotropy does not appear in the MSD; therefore, the statistical nature of diffusion must be examined using the fourth-order cumulant to detect anisotropy~\cite{han2006brownian}. When the particle orientation is hidden with no measurement noise, the MSD and fourth-order cumulant are given by 
\begin{align}
    &D_{xx}(t) = D_{yy}(t) = \bar{D} \label{2th}\\
    C_4(t) &= \frac{3\Delta D^2}{4D_\theta}\qty( t- \frac{1-\exp(-4D_\theta t)}{4D_\theta}).  \label{4th}
\end{align}

When the three diffusion coefficients are estimated using a fourth-order cumulant, two parameters must be estimated from a single equation, because $D_\theta$ and $\Delta D$ are coupled in Eq. ~[\ref{4th}]. In addition, in the $D_\theta \to \infty$ limit, the fourth-order cumulant vanishes regardless of time, and estimating $\Delta D$ and $D_\theta$ from the fourth-order cumulant is, in principle, impossible regardless of the length of each particle trajectory. Another drawback is that it does not use the full microscopic diffusion information contained in the time series.

\begin{figure}%[tbhp]
\centering
\includegraphics[width=1.0\linewidth]{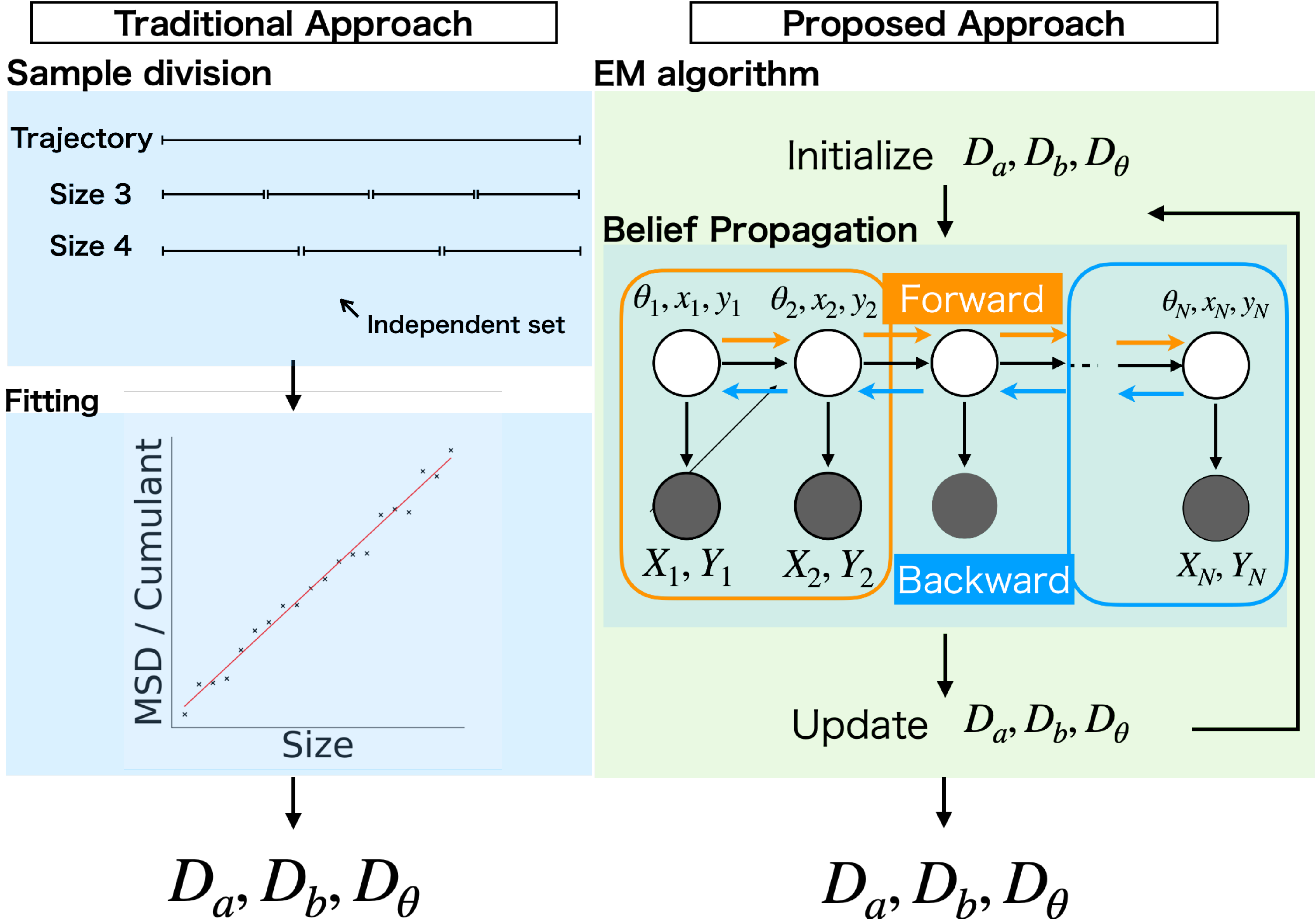}
\caption{Illustration of the traditional fitting approach and the proposed  
MLE approach. In the fitting approach, a single trajectory is segmented to multiple paths of varying lengths, 
from which the MSD and relevant cumulants are computed. In the MLE approach, on the other hand, 
given the initial values,  the diffusion coefficients are recursively updated by the EM algorithm based on 
relevant moments of latent variables with respect to the posterior distribution defined by 
the diffusion coefficients at the time. The moment assessment is efficiently performed 
by BP, which is implemented by particle filters.  }
\label{fig:ill}
\end{figure}

To address these limitations, we developed a novel method based on maximum likelihood estimation (MLE) for analyzing the mobility of anisotropic particles from the data of only single and short trajectories (Figure \ref{fig:ill}). The fitting method, which utilizes the mean values over time intervals, tends to lose information at higher moments. However, the MLE method efficiently exploits the information, including those with higher moments, from the data by maximizing the likelihood of the entire trajectory. The maximization is implemented using the expectation-maximization (EM) algorithm~\cite{dempster1977maximum} to incorporate latent variables. In the current system, belief propagation (BP)~\cite{pearl1988probabilistic} enables the algorithm to operate linearly with respect to the number of data points. However, in practice, BP is not feasible to perform analytically because it is expressed as a set of functional recurrence formulas. As a practical solution to this problem, we use particle-filter (sequential Monte Carlo) methods~\cite{kitagawa1996monte,liu1998sequential}.

The usefulness of our approach was validated by performing numerical studies and laboratory experiments using bacteria, quantum rods, and fluorescent spheres. 
We found that compared to conventional statistical methods widely used for estimating diffusion coefficients, our method can detect anisotropy more sensitively and estimate the diffusion coefficients of non-spherical tracer particles using shorter trajectories. Additionally, we analytically explored the detection limit of anisotropy, providing experimental guidelines regarding the length of the trajectories to obtain and the magnitude of the measurement noise to control.

\section{Results}

\subsection{Numerical studies}

\begin{figure}[tbhp]
\centering
\includegraphics[width=1.\linewidth]{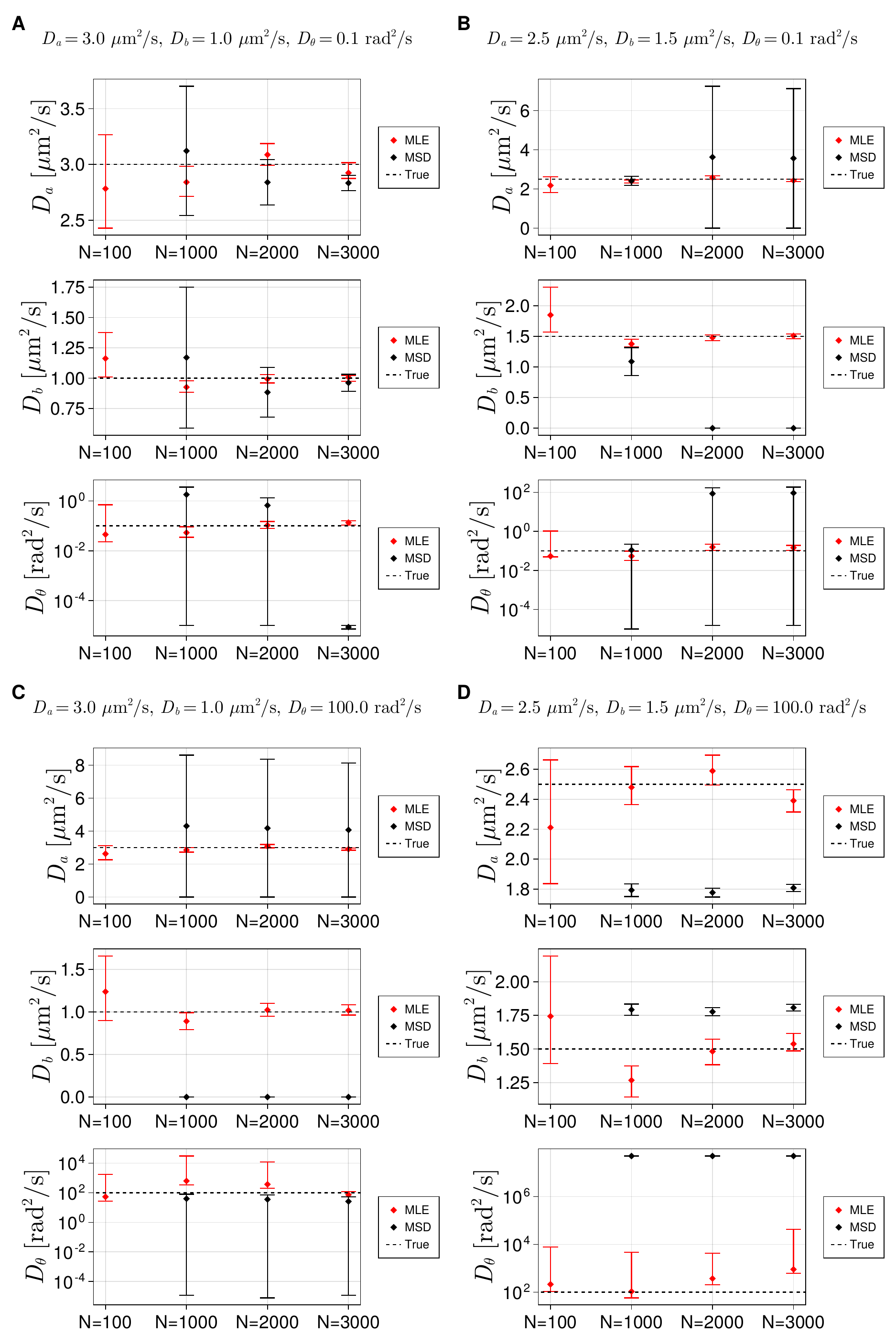}
\caption{Results of simulations performed using the MLE method compared to those performed using the fitting method. The results of the fitting method for $N=100$ are omitted because the error bars are too large and the estimation accuracy is obviously bad. Plots of $D_\theta$ are shown in log-scale.  The error bars represent the $\pm 1\sigma$ range obtained by Gaussian approximation of the log-likelihood function on either side of the maximum likelihood estimate.} (A-D) $\Delta t=\SI{0.01}{\second}, \epsilon=\SI{0.02}{\um}$.
\label{fig:sim}
\end{figure}

Figure \ref{fig:sim} shows the diffusion coefficients estimated from the trajectories generated by the numerical simulations with known diffusion coefficients. The simulations were conducted using four different parameter sets. In all cases, the MLE method outperformed conventional cumulant-based fitting. The fitting method achieved some degree of success in estimating \(D_a\) and \(D_b\) only when the rotational diffusion coefficient was small, that is \(\sqrt{2D_{\theta} \Delta t} < 1\) (Figure \ref{fig:sim} A,B). However, the MLE method succeeded in estimating both \(D_a\) and \(D_b\) with higher precision, even for shorter trajectories. Moreover, the MLE method could accurately estimate the correct order of magnitude for \(D_{\theta}\) even for trajectories as short as \(N=100\), which is impossible to achieve using the fitting method.

In more challenging scenarios where \(\sqrt{2D_{\theta} \Delta t} > 1\) (Figure \ref{fig:sim} C,D), the fitting method completely failed to estimate \(D_a\) and \(D_b\). In particular, Figure \ref{fig:sim} C shows an erroneous estimation of \(D_b = 0\), and Figure \ref{fig:sim} D represents an incorrect conclusion of \(D_a = D_b\). This failure is attributable to the fact that the fourth-order cumulant [\ref{4th}] vanishes asymptotically as \(D_{\theta}\) increases, thereby making the fitting process highly challenging. In particular, in Figure. \ref{fig:sim} D, distinguishing between isotropic and anisotropic diffusion becomes exceedingly difficult. Nonetheless, the MLE method successfully estimates \(D_a\) and \(D_b\) with high accuracy in both cases, and also successfully detects anisotropy.

\begin{figure}%[tbhp]
\centering
\includegraphics[width=1.\linewidth]{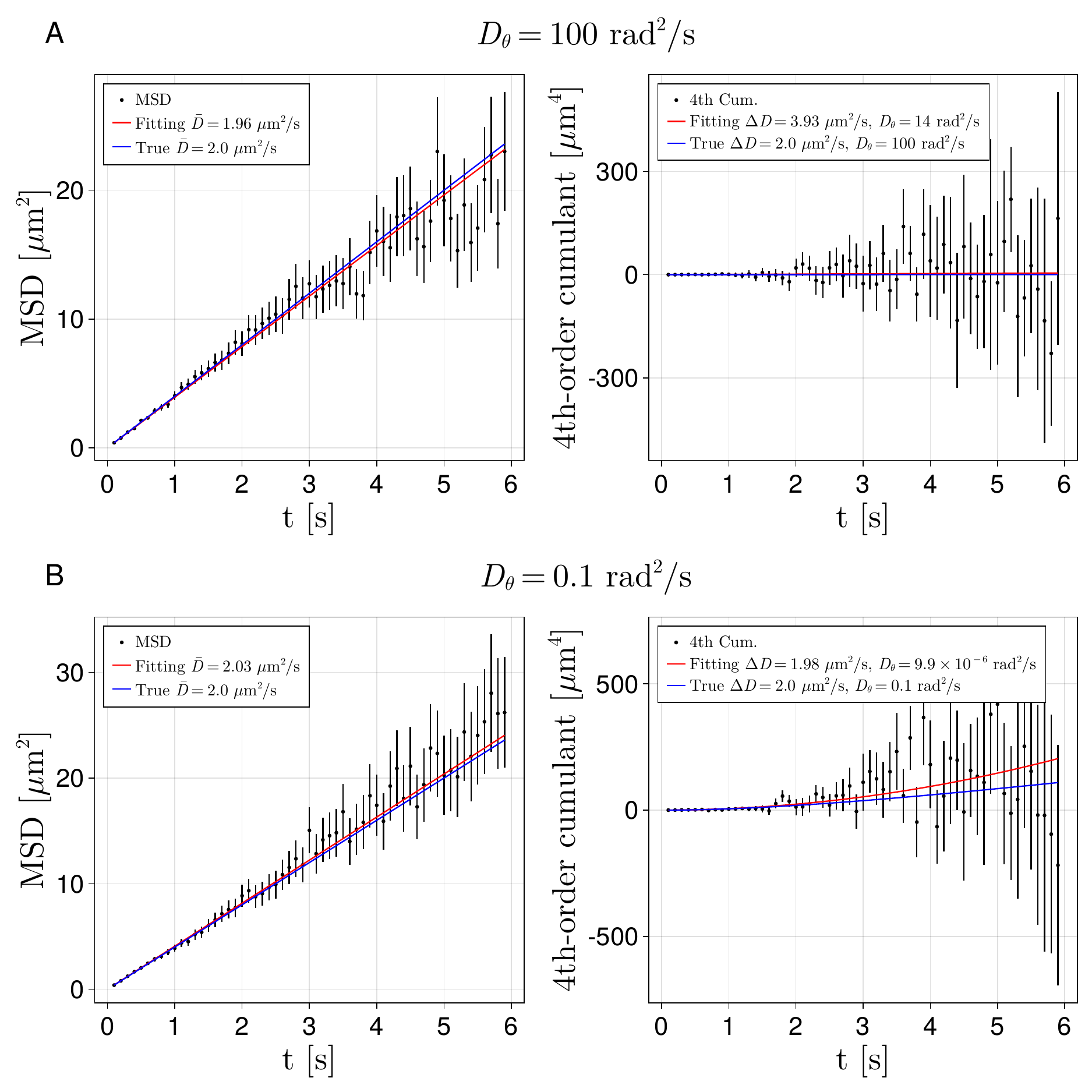}
\caption{
MSD and the fourth-cumulant estimated by fitting to simulation data. (A,B) $D_a=\SI{3.0}{\um^2 \per \second}$,$D_b=\SI{1.0}{\um^2 \per \second}$, $\Delta t=\SI{0.01}{\second}$, $\epsilon=\SI{0.02}{\um}$, $N=3000$. (A)$D_\theta=\SI{100.0}{\radian^2 \per \second}$. Error bars represent $\expval{\Delta x(t)^2 + \Delta y(t)^2}/\sqrt{2n}$. (B)$D_\theta=\SI{0.1}{\radian^2 \per \second}$. Error bars stand for $\sqrt{6} \expval{(\Delta x(t)^2 + \Delta y(t)^2)^2}/\sqrt{4n}$. 
For both cases, $n$ is the number of simulations (see Supplemental Material \cite{supp}).}
\label{fig:fitting}
\end{figure}

We found that the fourth-order cumulant is not useful for determining $\Delta D$ and $D_\theta$ because it exhibits significant statistical errors (Figure \ref{fig:fitting} in the large $t$ regimes). This leads to practical difficulties in accurately estimating $\Delta D$ and $D_\theta$. In particular, when $D_\theta$ is large, $C_4(t)$ does not respond to variations in $\Delta D$, which results in complete failure of the estimation, as shown in Figure \ref{fig:fitting} A. Moreover, as $D_\theta$ decreases, the cumulant method fails to predict even the correct order of $D_\theta$, as shown in Figure \ref{fig:fitting} B.

\begin{figure}%[tbhp]
\centering
\includegraphics[width=1.\linewidth]{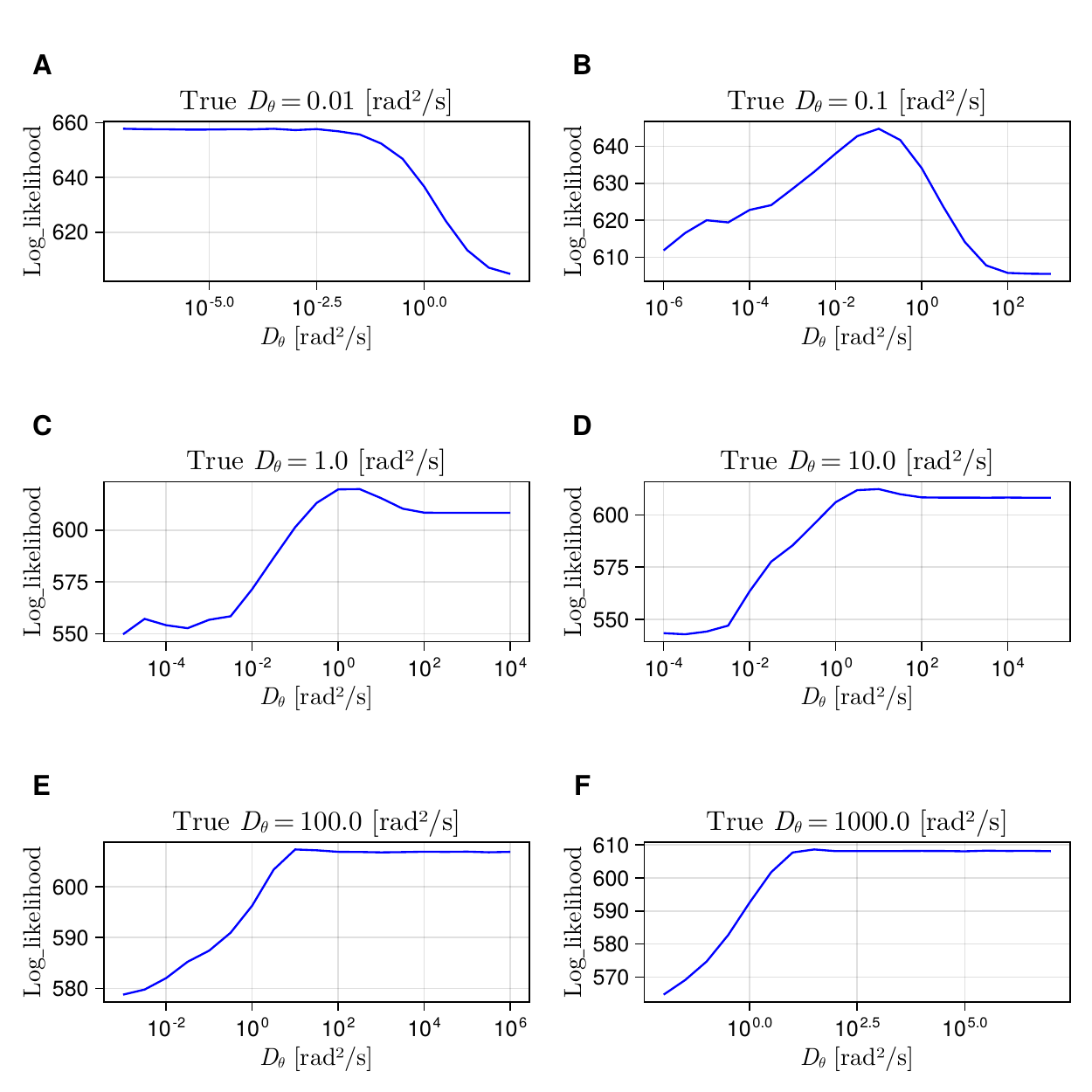}
\caption{Log-likelihood versus  $D_\theta$ when $D_a$ and $D_b$ are fixed to their true values. (A)  When the true value of $D_\theta$ is too small, distinguishing the influence of measurement noise from that of the rotational diffusion is difficult, making the true value indistinguishable from smaller values of $D_\theta$. (B,C) The correct order of $D_\theta$ can be estimated only when the true value of $D_\theta$ is moderate. (D-F) Conversely, when the true $D_\theta$ is too large, due to the $\pm\pi$ periodicity of the angle variables, distinguishing it from larger values is difficult. (A-F) $Da=\SI{2.0}{\um^2 / \second}$, $D_b=\SI{1.0}{\um^2 / \second}$, $\Delta t=\SI{0.01}{\second}$, and $\epsilon=\SI{0.02}{\um}$.}
\label{fig:log_like}
\end{figure}

As verified in Figures \ref{fig:sim} A and B, our method can estimate not only the translational diffusion coefficients $D_a$ and $D_b$, but also the rotational diffusion coefficient $D_\theta$, even though the angle $\theta$ is hidden. However, Figures \ref{fig:sim} C and D also show that if $\sqrt{2D_\theta \Delta t} > 1$, the estimation error of $D_\theta$ increases, because the angles are determined almost independently at each time step. To clarify this phenomenon further, Figure \ref{fig:log_like} shows the landscape of the likelihood function for the true value of $D_\theta$.  When the rotational diffusion coefficient is too small (\(\sqrt{2 D_\theta \Delta t} \ll 1\)), the likelihood function lacks extrema, making it impossible to estimate its order (Figure \ref{fig:log_like} A). This can be attributed to the effect of measurement noise, which becomes significant owing to the absence of angular observations. Conversely, when the rotational diffusion coefficient is large (\(\sqrt{2 D_\theta \Delta t} > 1\)), the $\pm\pi$ periodicity of the angles causes the likelihood function to lack extrema (Figure \ref{fig:log_like} E,F). Therefore, as for the angular diffusion, our method may be effective only when 
the true $D_\theta$ has a moderate value. 
Nevertheless, even when accurate estimation of the order of the rotational diffusion coefficient is not possible, the translational diffusion coefficients can still be accurately estimated. Finally, one point that requires consideration is that this is not a discussion of the practical performance of the estimation algorithm, but instead concerns the general estimability with respect to the likelihood function.

\subsection{Application to biological systems}
\begin{figure}%[tbhp]
\centering
\includegraphics[width=1.\linewidth]{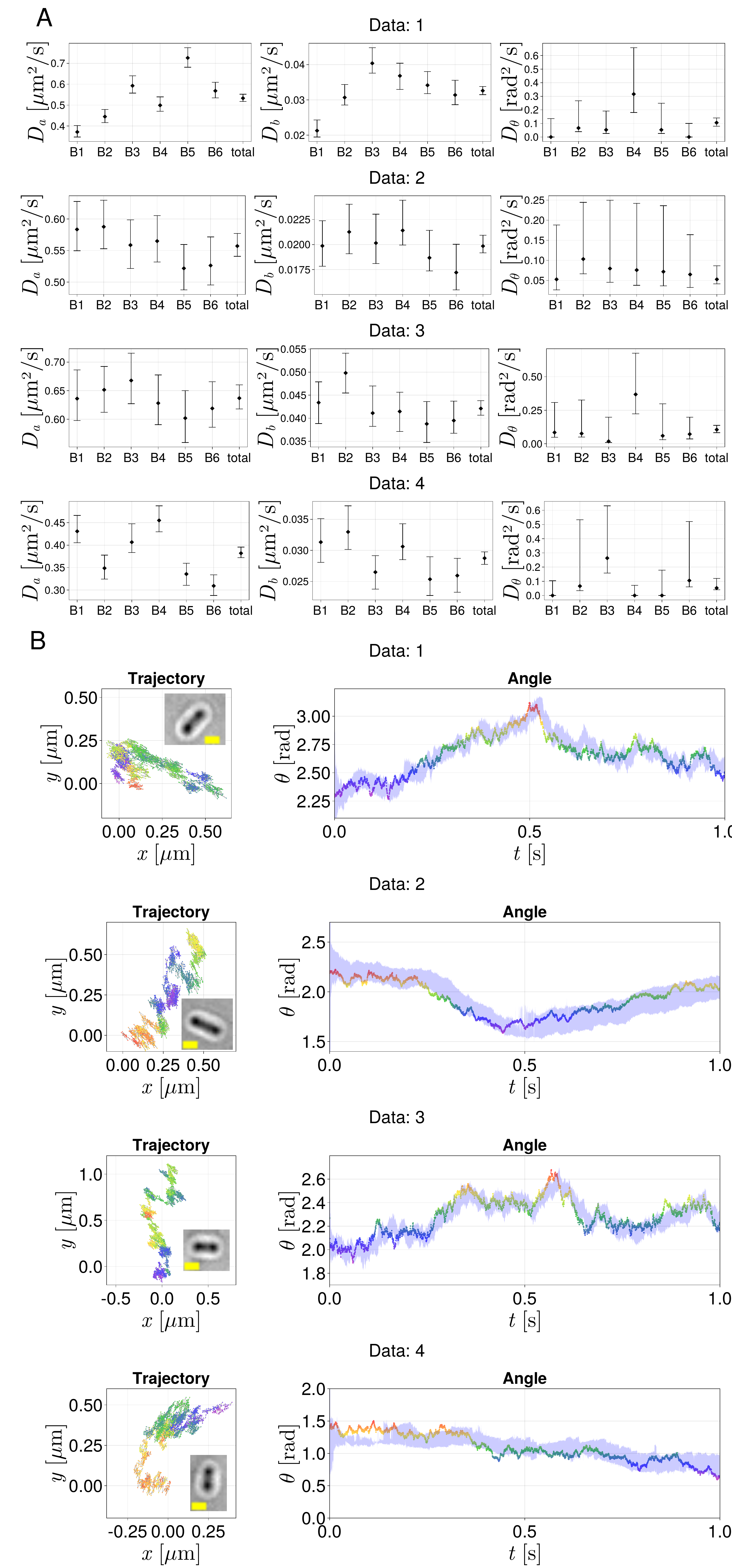}
\caption{Results obtained by the MLE method for four diffusion trajectories of bacteria. (A) The results obtained by dividing the $N=3000$ trajectory into six subsets and estimating the diffusion coefficients in each $N=500$ block are compared to the result obtained from the whole $N=3000$ data. The error bars represent the $\pm 1\sigma$ range obtained by Gaussian approximation of the log-likelihood function on either side of the maximum likelihood estimate.} (B) Actual trajectories (left column) and estimated angles (right column).  The inset in each trajectory data shows a snapshot of a bacterium with a yellow scale bar ($\SI{1}{\um}$). The angles at each time point are color-coded. The blue bands indicate the $90\%$ confidence intervals of the estimated angles. (A,B) $\Delta t = \SI{0.33}{\ms}$ and $\epsilon=\SI{0.00395}{\um}$.
\label{fig:angle}
\end{figure}

The numerical results in the previous section confirm that our MLE-based method significantly outperforms conventional cumulant-based methods for a wide range of diffusion coefficients. To verify the utility of this method for real biological samples, we estimated the diffusion coefficients of micron-sized bacterial cells  in quasi-two-dimensional suspensions confined between two parallel glass walls (Figure \ref{fig:angle} A). First, we extracted trajectory segments of \(N=3000\) length from four independent datasets, partitioned them into six \(N=500\) subsets, and estimated the diffusion coefficients for each. The results were compared with the estimates using the full \(N=3000\) dataset. Interestingly, for Data2 and Data3, the estimated diffusion coefficients did not vary significantly across the subsets. In contrast, for Data1 and Data4, the translational diffusion coefficients appeared to particularly differ across subsets. While this may be attributed to fluctuations in the distance between the bacteria and chamber walls~ \cite{bitter2017interfacial}, the salient point is that the MLE method can detect such time-dependent variations in the diffusion coefficients even over short intervals. This indicates the potential of the method to identify nonstationary particle behaviors during diffusion from short-length trajectories.

Micron-sized bacteria allow for angle observations, which enable the verification of agreement between the estimated angles and actual observations. Figure \ref{fig:angle} B compares the estimated angular distributions \(p(\theta \mid \vb{X},\vb{Y}, \boldsymbol{D}^\ast)\) with the actual observations, where \(\boldsymbol{D}^\ast\) represents the estimated diffusion coefficients. Although we did not utilize any angular observation information to derive \(p(\boldsymbol{\theta} \mid \vb{X},\vb{Y}, \boldsymbol{D}^\ast)\), Figure \ref{fig:angle} B shows a good agreement between the estimated and observed bacterial angles. Thus, even when the particles are too small for directional detection, the directional behavior can still be estimated with high precision from noise-added trajectories alone as a simultaneous outcome of diffusion coefficient estimation.

\subsection{Application to nanosized anisotropic particles}
\begin{figure}%[tbhp]
\centering
\includegraphics[width=1.\linewidth]{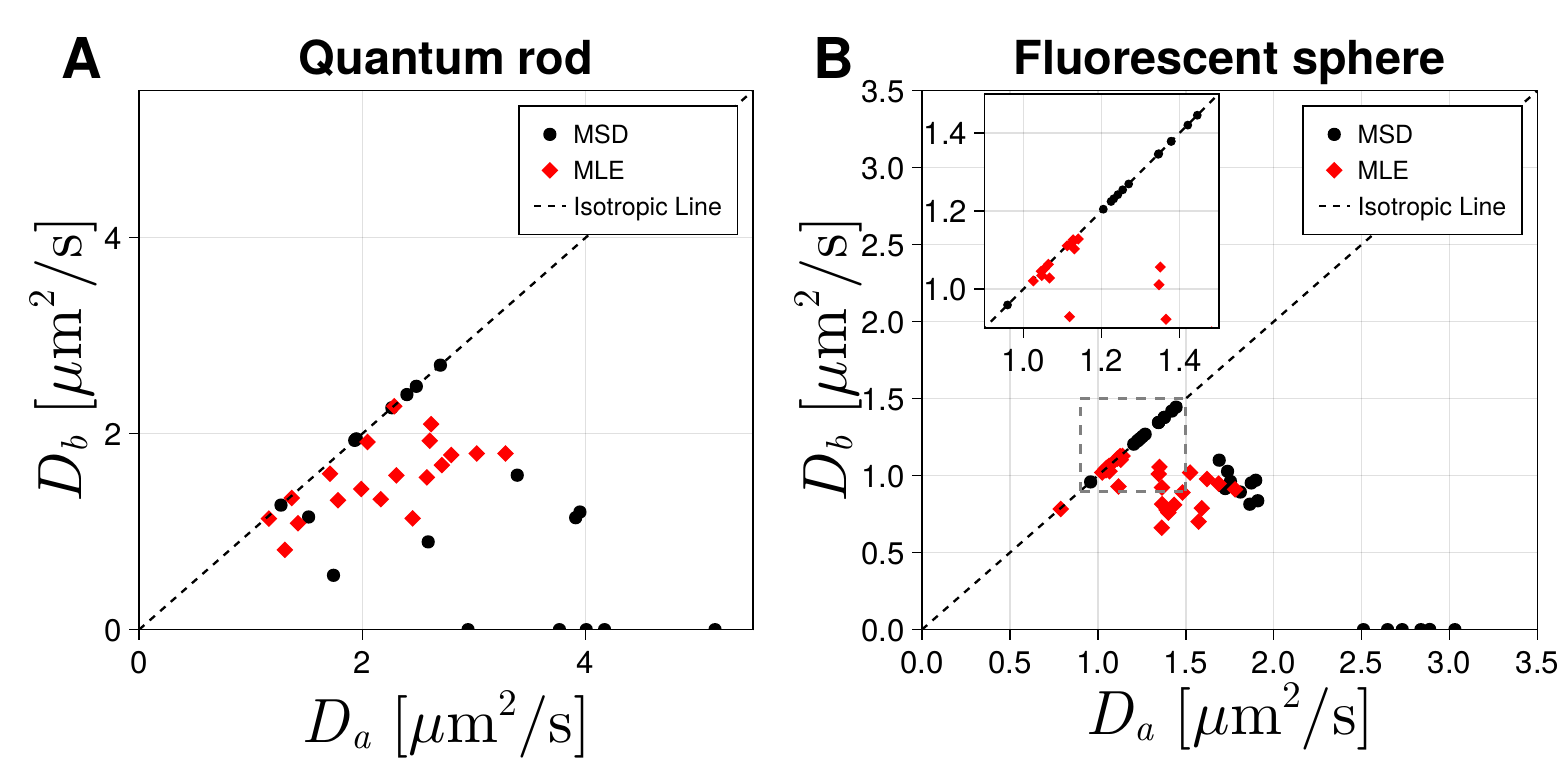}
\caption{Results obtained by the MLE method for the diffusion trajectories of (A) quantum rods and (B) fluorescent spheres. The nominal major and minor axis lengths of the quantum rod are 28.4$\pm$3.0 and 4.6$\pm$0.7 nm, respectively. The nominal diameter of the fluorescent spheres is 200 nm. (A,B) $N=1000, \Delta t = \SI{0.01}{\second}$ and $\epsilon=\SI{0.02}{\um}$.}
\label{fig:rod}
\end{figure}

Owing to the slow rotational diffusion and pronounced anisotropy of micron-sized bacteria (Figure \ref{fig:angle} B), detecting diffusion anisotropy in these bacteria is relatively easy. To further validate the effectiveness of our approach, we examined  more challenging nanosized systems, namely the diffusion of quantum rods and fluorescent spheres, which exhibit lesser anisotropy and faster rotational diffusion. Figure \ref{fig:rod} A shows the estimated translational diffusion coefficients for $19$ independent trajectories of the quantum rods. The conventional fitting approach tended to either incorrectly infer isotropic behavior (\(D_a = D_b\)) or produce extreme parameters, such as \(D_b = 0\). This can be attributed to a failure in fitting the fourth-order cumulant, which yielded erroneous estimations that fall into the extremities of the parameter space, such as \(\Delta D = 0\) or \(\Delta D = D_a\). Conversely, the estimates obtained using our method eliminate extreme solutions such as \(D_b = 0\) and reduce the fraction of trajectories classified as isotropic. This suggests that our method can detect anisotropy with high accuracy even in trajectories exhibiting very low non-Gaussianity, where the fourth-order cumulant approaches zero. However, some data were still classified incorrectly as isotropic.

\subsection{Application to nanosized isotropic particles}
\begin{figure*}%[tbhp]
\centering
\includegraphics[width=\textwidth]{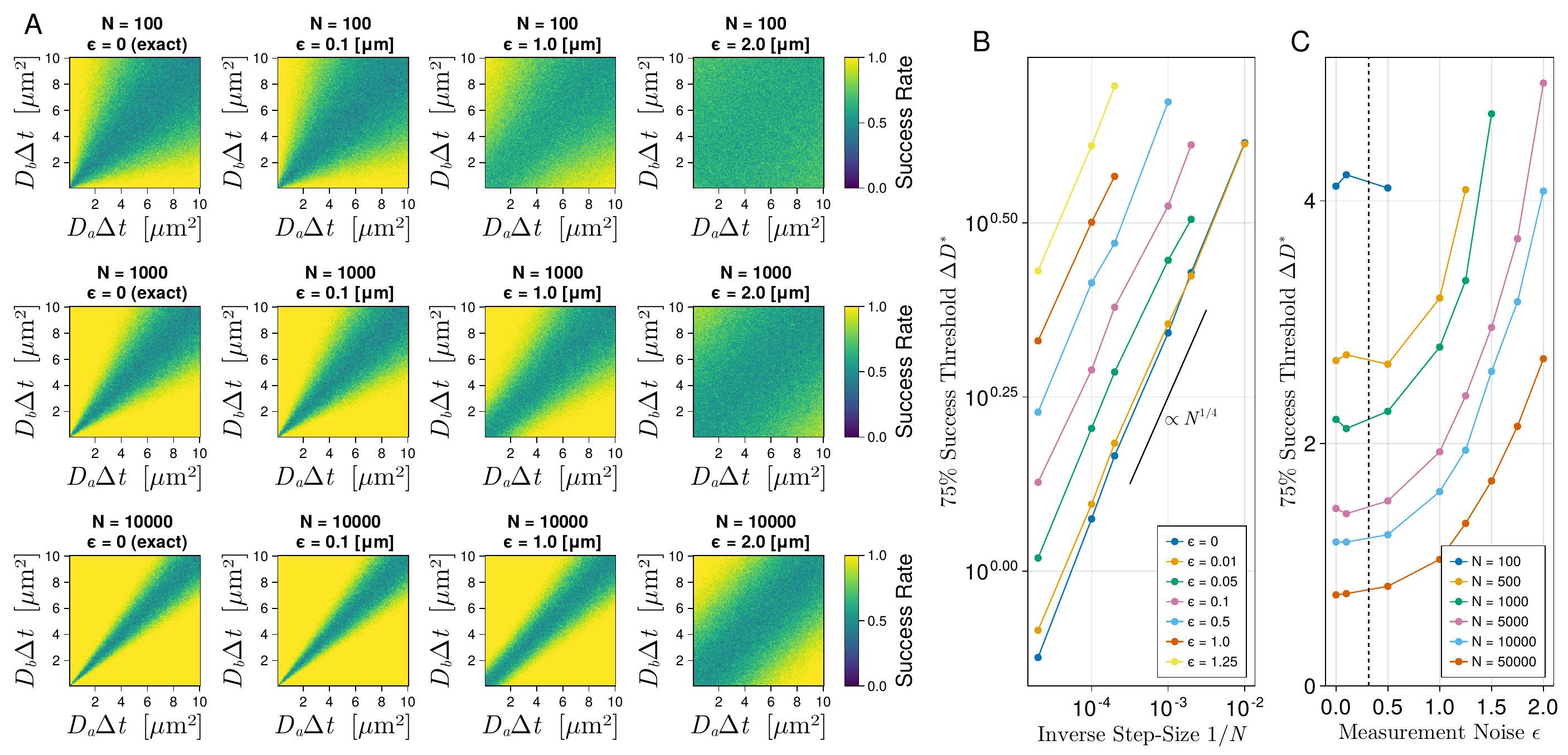}
\caption{Influence of trajectory length $N$ and magnitude of measurement noise $\epsilon$ on the success rate of anisotropy detection. 
%$\Delta D^\ast$ is a threshold at which the success probability exceeds $75\%$ for the first time. 
(A) Heat map of the success rate on the $D_a-D_b$ plane.  (B) We define $\Delta D^\ast$ 
as the width of the interval on $D_a\Delta t + D_b\Delta t = \SI{10.0}{\um^2 }$ on which the success rate is smaller than $75\%$. 
The plots indicate that $\Delta D^\ast$ vanishes as ${\mathcal O}(N^{-1/4})$ as $N$ tends to infinity. (C) When the noise magnitude 
$\epsilon$ is smaller than the dashed line (corresponding to $\epsilon$ where the signal-to-noise ratio is 1), 
the measurement noise has little impact on the success rate. However, for $\epsilon$ larger than the dashed line, the noise magnitude significantly influences the success rate. (B-C) 
Plots are obtained for $\bar{D} = \SI{10.0}{\um^2 / \second}$ and $\Delta t = \SI{0.01}{\second}$.
% the plots illustrate the variation of $\Delta D^\ast$, which is the smallest value that first exceeds a $75\%$ success rate for detecting anisotropy.
}
\label{fig:vs_D}
\end{figure*}

Our method was originally formulated to detect diffusion anisotropy, but can also handle isotropic diffusion in the limit case of \(D_a = D_b\). To confirm the detection performance for isotropic diffusion, we used the trajectories of fluorescent spheres to estimate the diffusion coefficients (Figure \ref{fig:rod} B). Almost half of the 28 trajectories were classified as isotropic, which is consistent with the theoretical probability of detecting diffusion isotropy, the details of which are discussed in the next section (Figure \ref{fig:vs_D} A). 

\subsection{Stability Analysis of the EM algorithm}
The results in Figure \ref{fig:rod} A(B) suggest that several trajectories can lead to incorrect isotropic(anisotropic) estimation outcomes, even using our MLE method when $D_\theta$ is sufficiently large. To investigate the reason for this, we theoretically analyzed the dynamic behavior of the EM algorithm. 
Because of the symmetry between $D_a$ and $D_b$, an isotropic fixed point always exists in the EM algorithm. 
%In the updated equations of the EM algorithm, when $D_\theta$ is sufficiently large, an isotropic solution becomes a fixed point. 
To examine whether an isotropic fixed point is preferable, 
we linearized the EM algorithm around this point and examined the resulting stability matrix. 
%We calculated and analyzed the eigenvalues of the linearized evolution matrix of the system at this fixed point. 
This yielded the eigenvalue $q$ corresponding to the eigenvector that breaks the isotropy as follows: 
\begin{align}
  q = 1+  \qty(\frac{D_a-D_b}{D_a+D_b})^2 \qty(\frac{1}{e^{4D_\theta\Delta t}-1} +\frac{1}{2}) \label{q_main}
\end{align}
in the $N\to\infty$ limit and $\epsilon=0$. The derivation is provided in the Supplemental Material \cite{supp}. This result implies that it always exceeds $1$ as long as $D_a  \neq D_b$; that is, if the particle is anisotropic, the isotropic fixed point is unstable. Therefore, anisotropic particle properties can always be detected if an infinitely long trajectory is observable without measurement noise even if $D_\theta\to \infty$. The reason for this counterintuitive conclusion is that in 
Eq.~[\ref{4th}], $D_\theta$ and $\Delta D$ are completely coupled, such that the fourth-order cumulant converges to zero when $D_\theta$ is infinite, whereas in Eq. ~[\ref{q_main}], there exists a term in which $\Delta D$ and $D_\theta$ are decoupled by the term $+1/2$. This decoupling term remains even if $D_\theta\to \infty$, enabling the detection of diffusion anisotropy. 

Nevertheless, in realistic situations, the length of the observable trajectory is finite and measurement noise is inevitable. Under these conditions, the results described by Eq. ~[\ref{q_main}] no longer hold, and the eigenvalues may dip below unity, depending on the statistical fluctuations of the sampled trajectory. In other words, even if the particle is non-spherical, the observed trajectory may display isotropic characteristics by chance, making the isotropic solution locally optimal. (Note that this is actually globally optimal in most cases; see Supplemental Material \cite{supp}).  
Figure \ref{fig:vs_D} A illustrates the probability of observing a trajectory with an eigenvalue $q$ greater than $1$ and summarizes the dependency of this probability on $N$, $\epsilon$, $D_a$, and $D_b$.
As expected, this figure shows that the larger the difference in the diffusion coefficient $\Delta D$, the higher the probability of successful anisotropy detection.

We also introduced $\Delta D^\ast$ as the minimum $\Delta D$ value with a $75\%$ probability of an eigenvalue exceeding $1$ on the line of %$\bar{D}=\SI{10.0}{\um^2 / \second}$, 
$D_a\Delta t + D_b\Delta t = \SI{10.0}{\um^2 }$,
which serves as the detection threshold for anisotropy. Using $\Delta D^\ast$, we first revealed that the probability of the eigenvalues exceeding $1$ asymptotically approaches $100\%$ within the limit of an infinitely large number of observations (Figure \ref{fig:vs_D}). This result indicates that our method reliably detects anisotropy even in the presence of observation noise within the limit of $N\to \infty$.
Second, we found that the threshold $\Delta D^\ast$ decreased with $\mathcal{O}(N^{-1/4})$ with respect to the trajectory length $N$ and that the order is almost independent of noise (Figure \ref{fig:vs_D} B). This result demonstrates that the estimation accuracy depends moderately on the trajectory length.
Finally, we verified that the amplitude of the measurement noise has little effect on the probability of detecting anisotropy when $\epsilon< \epsilon^\ast$, where $\epsilon^\ast$ corresponds to a diffusion signal-to-noise ratio of $1$, that is, $\epsilon^\ast = \sqrt{2\bar{D}\Delta t}$. In contrast, we found that the amplitude of the measurement noise has a substantial effect on the probability of detecting anisotropy when the magnitude of noise exceeds the typical magnitude of diffusion ($\epsilon > \epsilon^\ast$) (Figure \ref{fig:vs_D} C).

 These results can be used to assess the number of observations and noise level required to reliably detect anisotropy. 
Specifically, by performing the numerical simulations used to create Figure \ref{fig:vs_D}, we can evaluate $N$ and $\epsilon$ required to achieve an acceptable level of success probability of anisotropy detection for each estimation result. 
Furthermore, Figure \ref{fig:vs_D} also offers qualitative guidance for improving the performance of anisotropy detection: Increasing the number of observations is more effective than reducing the measurement noise when the noise is smaller than $\epsilon^\ast$. We remark that this result is not specific to the EM algorithm because it relates to the landscape of the likelihood function and is algorithm-independent as long as any type of stochastic approach is used for the estimation. Details of the theoretical analysis of the probability of detecting anisotropy are described in the Supplemental Material \cite{supp}.

\begin{figure}%[tbhp]
\centering
\includegraphics[width=1.\linewidth]{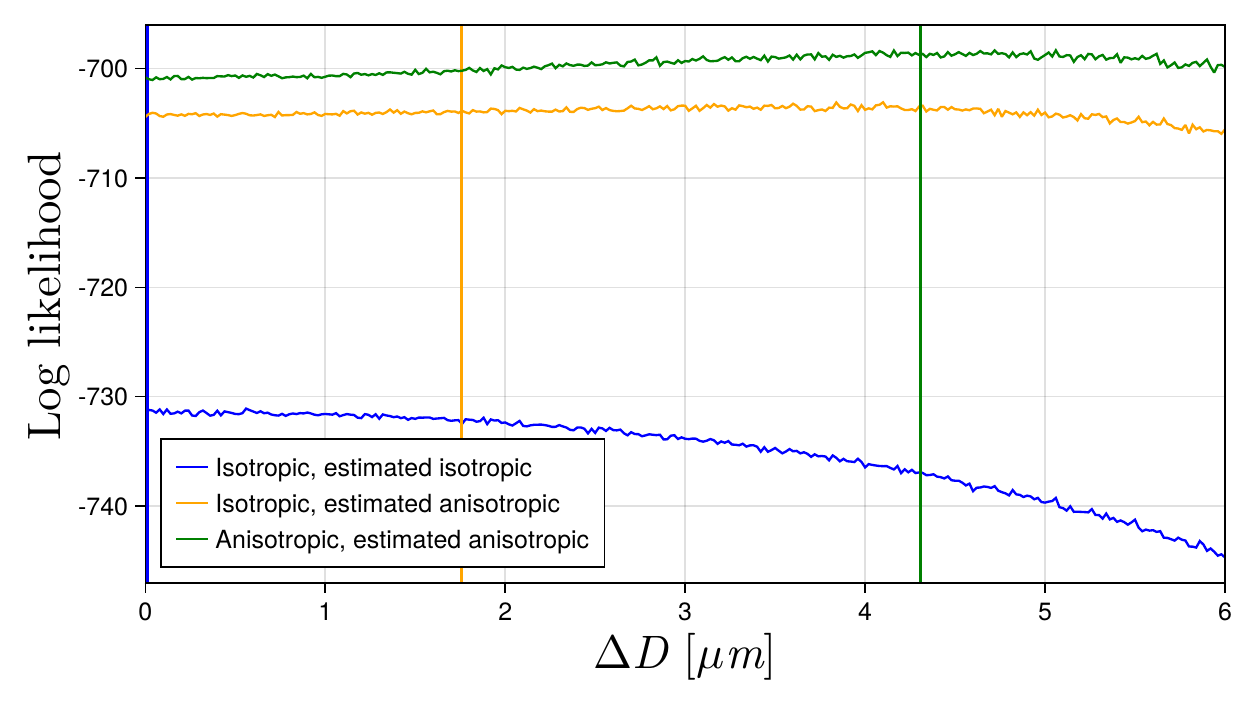}
\caption{ Typical profiles of the log-likelihood functions in three cases: when a particle is isotropic but is incorrectly estimated to be anisotropic, when it is correctly estimated to be isotropic, and when a particle is anisotropic and correctly estimated to be anisotropic. The vertical lines represent the maximum likelihood estimates in each case.  $N=1000, \Delta t = \SI{0.01}{\second}$ and $\epsilon=\SI{0.1}{\um}$. In isotropic case $D_a =D_b = \SI{5.0}{\um^2 / \second}, D_\theta = \SI{100}{\radian^2 / \second} $ and in anisotropic case $D_a = \SI{7.0}{\um^2 / \second}, D_b = \SI{3.0}{\um^2 / \second}, D_\theta = \SI{100}{\radian^2 / \second}  $.}
\label{fig:Ddiff}
\end{figure}

From Figure \ref{fig:rod}, we can see that even if the particles are isotropic, anisotropy is falsely detected with a certain probability and vice versa.
This is consistent with theoretical analysis shown in Figure \ref{fig:vs_D}, which poses significant challenges regarding the reliability of the analytical results of experimental data.
One possible solution is to use Bayesian inference, which enhances estimation accuracy by introducing a prior distribution to the estimated parameters when prior knowledge is available.
Another approach is to examine the profile of the likelihood function. Figure \ref{fig:Ddiff} illustrates typical profiles of the likelihood function for three scenarios.  
One scenario depicts an isotropic particle being mistakenly estimated as anisotropic. 
The other two scenarios refer to cases where isotropic/anisotropic particles are correctly estimated to be isotropic/anisotropic, respectively.
This figure shows that when an isotropic particle is incorrectly estimated to be anisotropic, the likelihood function is much flatter than in the other two correctly estimated cases. 
This means that the estimation result is much less reliable and one should be very careful when accepting it.

\section{Discussion}
The main subject of our research is the estimation of diffusion coefficients and the detection of diffusion anisotropy based on an anisotropic diffusion model that does not presuppose angular observations. To this end, we propose a novel MLE method that uses an EM algorithm. We successfully addressed a complicated system containing unobservables, that is, angles, by employing BP within the EM framework and evaluating the posterior distribution using particle filtering. 

The diffusion model presented in this study is highly universal, relevant to many real systems, and has promising broad applications across various fields. For example, our method provides a new framework for estimating the diffusion parameters of a system, even for time-series data in economics where it is difficult to prepare homogeneous statistical ensembles, or for data generated from simulations with very high computational costs.
To the best of our knowledge, our method is the first attempt to utilize BP within the MLE for diffusion estimation problems, which opens the door for its use in more complex diffusive systems, including those with external fields~\cite{dessard2024cytoplasmic}, confined geometries~\cite{godin2017single}, and hydrodynamic background driven by biological activities~\cite{peng2016diffusion}. 

%Furthermore, by using 
Using numerical studies and experimental tests, we demonstrated that the MLE method outperforms conventional methods based on MSD and fourth-order cumulants in various parameter regimes. Conventional methods often failed to estimate diffusion coefficients, even with \(N > 3000\), whereas our MLE method was successful with as few as \(N = 100\) observations. This ensured that our method is applicable to realistic situations in which the macroscopic non-Gaussianity is almost entirely masked owing to high rotational diffusion coefficients.

Moreover, our theoretical analysis revealed that the probability of anisotropy detectability depends on the trajectory length \(N\) and measurement noise \(\epsilon\). First, we showed that the success of the MLE method lies in its ability to decouple the dependencies between \(\Delta D\) and \(D_\theta\), which are strongly coupled in the fourth-order cumulants. Second, we found that in the limit where \(D_\theta\) approaches infinity, anisotropy is theoretically always detectable for infinite trajectory lengths, whereas the finiteness of the lengths can sometimes lead to an isotropic solution as the optimal solution. This indicates a theoretical limit for anisotropy detection, indicating the need for improvements to the measurement precision rather than algorithmic enhancements. Thus, this study provides both qualitative and quantitative guidelines for increasing the measurement accuracy. Finally, a comparison of our outcome with that of recent research~\cite{Fukuda2023-rs} on binary classification between anisotropic and isotropic diffusion is of interest. Because the previous study employed deep learning and focused solely on binary classification, the underlying estimation principle was largely a black box, and quantitative estimation of the anisotropic diffusion coefficients remained impossible. In contrast, our results provide a theoretical basis for the limitations of the success probability of estimation, in addition to quantitative and reliable detection.

\section{method}
\subsection{Overview of Estimation Algorithm}
To estimate the diffusion coefficients and detect their anisotropy, instead of the conventional cumulant-based methods, we employed MLE, which has been shown to be useful for estimating diffusion coefficients in much simpler models~\cite{Shuang2013-tg, Makkai2023-dx, Lin2021-ne, Bullerjahn2021-zz} or some specific cases~\cite{Haas2013-om, Koo2015-sn}. 

First, Eq. ~[\ref{Lan}] was discretized using Ito's method to convert the diffusion process into a set of stochastic equations as follows:
\begin{align}
    (x_{i+1}-x_{i}, y_{i+1}-y_i) &\sim \mathbf{Norm}(\vb{0}, \boldsymbol{\Sigma}_i\Delta t), \\
        \theta_{i+1}-\theta_i &\sim \mathbf{Norm}(0, 2 D_\theta\Delta t ), \\
        X_i &\sim \mathbf{Norm}(x_i, \epsilon^2 ), \\
        Y_i &\sim \mathbf{Norm}(y_i, \epsilon^2 ),
\end{align}
where $\boldsymbol{\Sigma}_i=\boldsymbol{\Sigma}(\theta_i)$. From these equations, we derived the following expression for the likelihood function:
\begin{align}
    &p(\vb{X}, \vb{Y} \mid \vb{D} )  \propto \small \int d\vb{x}d\vb{y} d\boldsymbol{\theta} \notag\\
    ~~~~~~~~~~~~~~&\exp(-\sum_i \qty{\frac{1}{4\Delta t}\qty(\frac{u_i^2}{D_a}+\frac{v_i^2}{D_b} + \frac{\Delta \theta_i^2}{D_\theta})}) \notag\\
    ~~~~~~~~~~~~~~&\times\exp( \frac{(x_i-X_i)^2+(y_i-Y_i)}{2\epsilon^2})\label{like}
\end{align}
where $\vb{D} = (D_a, D_b, D_\theta)$ and 
\begin{align}
  u_i &= \cos{\theta_i}\Delta x_i + \sin{\theta_i}\Delta y_i\\
  v_i &= \sin{\theta_i}\Delta x_i -\cos{\theta_i}\Delta y_i .
\end{align}

Our objective was to estimate $\vb{D}$ using Eq.~[\ref{like}]. This could be performed using the Markov chain Monte Carlo method. 
However, this approach is not feasible because the transition probabilities that converge to $p(\vb{D} \mid \vb{X}, \vb{Y}) \propto p(\vb{X}, \vb{Y} \mid
\vb{D})$ generally scale exponentially with respect to $N$, which prevents accurate estimation of $\vb{D}$ and produces huge statistical fluctuations in sampling. 
Therefore, we used the MLE method instead, which is free from large statistical fluctuations and can be performed using the EM algorithm with $\mathcal{O}(N)$ computational cost per update, as shown below. 

The EM algorithm that iterates
\begin{align}
  D_a^{k+1} &= \frac{1}{2T} \expval{\sum_{i=1}^{N-1}  (\cos{\theta_i}\Delta x_i + \sin{\theta_i} \Delta y_i)^2}_{\vb{D}^k}  \label{Da}\\
  D_b^{k+1} &= \frac{1}{2T} \expval{\sum_{i=1}^{N-1} (\sin{\theta_i}\Delta x_i -\cos{\theta_i}\Delta y_i)^2}_{\vb{D}^k}   \label{Db}\\
  D_\theta^{k+1} &= \frac{1}{2T} \expval{\sum_{i=1}^{N-1} \Delta \theta_i^2}_{\vb{D}^k}  \label{Dtheta}
\end{align}
with the appropriate initial conditions is guaranteed to converge to a local maximum of %the likelihood 
Eq.~[\ref{like}], 
 where $k$ counts the number of iterations, $\vb{D}^k = (D_a^k, D_b^k, D_\theta^k)$,
 $T=N\Delta t$, and $\Delta x_i = x_{i+1} - x_{i},\Delta y_i = x_{i+1} - y_{i},\Delta \theta_i = \theta_{i+1} - \theta_{i} $.  
The expectation $\expval{\cdots}_{\vb{D}^k}$ is considered under the posterior distribution  
 $p(\vb{x}, \vb{y}, \boldsymbol{\theta} \mid \vb{X}, \vb{Y}, \vb{D}^k)$. 

The subsequent step involves computing the expectations on the right-hand side of Eqs. [\ref{Da}], [\ref{Db}], and [\ref{Dtheta}]. 
We perform this using BP \cite{pearl1988probabilistic}, 
which is an efficient algorithm that is applicable to probabilistic models defined over cycle-free graphs.
In the current system, BP propagates auxiliary distributions, termed ``messages'', as
\begin{align}
    \nu_{i} (\vb{z}_{i}) &\propto \int d\vb{z}_{i-1}~ \nu_{i-1}(\vb{z}_{i-1})  p(\vb{z}_i \mid \vb{z}_{i-1})  p(x_i, y_i \mid X_i, Y_i)  \label{forward}\\ 
    \mu_{i} (\vb{z}_{i}) &\propto \int d\vb{z}_{i+1}~ \mu_{i+1}(\vb{z}_{i+1})  p(\vb{z}_i \mid \vb{z}_{i+1})  p(x_i, y_i \mid X_i, Y_i) \label{backward}
\end{align}

along a chain that corresponds to a sequence of latent variables $\vb{z}_i = (x_i, y_i, \theta_i)$ ($i=1,\ldots, N$)
in the forward and backward directions.
Once messages $\nu_{i} (\vb{z}_{i})$ and $\mu_{i} (\vb{z}_{i})$ have been computed
for $i=1,\ldots,N$, the joint posterior distribution can be assessed as
 \begin{align}
    p(\vb{z}_{i}, \vb{z}_{i+1}  \mid  \vb{X}, \vb{Y}, \vb{D}) \propto \nu_i(\vb{z}_i)p(\vb{z}_{i+1} \mid \vb{z}_{i})\mu_{i+1}(\vb{z}_{i+1}), \label{joint}
\end{align}

which makes it possible to efficiently compute the right side of Eqs. [\ref{Da}], [\ref{Db}], and [\ref{Dtheta}].

The cost for computing Eqs. [\ref{forward}] and [\ref{backward}] for $i=1,\ldots,N$ scales only linearly with respect to $N$. 
However, analytically performing the functional update of the BP remains challenging. To overcome this problem, 
we used sampling methods known as particle filters or sequential Monte Carlo methods \cite{kitagawa1996monte,liu1998sequential}. This approach uses a Monte Carlo approximation of the distribution, preserving it as a population of ``particles''. Although this induces some Monte Carlo errors and compromises exactness, augmenting the number of particles can lead to improved approximations. Particle degeneracy is an inherent issue in particle filters; however, in our model, adept interweaving of the diffusion-propagation step with the observation-resampling step effectively minimized the problems resulting from degeneracy. The detailed algorithm for preventing degeneration is described in Supplemental Material \cite{supp}.

Calculating the joint distribution [\ref{joint}] is inherently more challenging than calculating its marginalized counterpart \(p(\vb{z}_i \mid \vb{X}, \vb{Y}, \vb{D})\). 
%This arises because, while the distribution of angle \(\theta_i\) exhibits a \(2\pi\) periodicity, the distribution of the angle difference, \(\Delta \theta_i\), does not. 
This occurred because the absence of measured data meant that  
$\theta_i$ and $\theta_{i+1}$ sampled from $\nu_i(\vb{z}_i)$ and $\mu_{i+1}(\vb{z}_{i+1})$ were almost uncorrelated, yielding large statistical fluctuations when assessing Eq. [\ref{joint}] using naive particle filters. 
We have addressed the techniques to mitigate this challenge in the Supplemental Material \cite{supp}. In particular, \(D_\theta\) empirically showed extremely slow convergence in the EM algorithm, with convergent solutions often not reaching satisfactory approximations. Thus, by leveraging the efficiency of the particle filter in the likelihood function computation, we exclusively employed surrogate-based optimization \cite{Urquhart2020-bq} to determine the likelihood function when calculating the maximum likelihood value for \(D_\theta\). This algorithm allowed for approximate optimization with high precision, even when the objective function exhibited Monte Carlo fluctuations. Finally, an efficient method for computing the likelihood function using a particle filter is described in the Supplemental Material \cite{supp}.
ta
\subsection{Experiments}
The details of the sample preparation, microscopy, and image analysis are described in the Supplemental Material \cite{supp}.

\begin{acknowledgments}
This work was supported by JSPS KAKENHI grants Grant-in-Aid for Scientific Research on Innovative Areas, "Information Physics of Living Matters" (\#19H05794, \#19H05795) to Y.O. and Grant-in-Aid for Transformative Research Areas (A), "Foundation of Machine Learning Physics" (\#22H05117) to Y.K.
by the Japan Science and Technology Agency (JST; JPMJCR20E2, JPMJMS2025-14) to Y.O. (JPMJCR20E5, JPMJMS2022-14) to M.K., and 
(JPMJCR1912) to Y. K. 
\end{acknowledgments}

\bibliography{sorsamp.bib}% Produces the bibliography via BibTeX.

\clearpage
\onecolumngrid
\appendix
\pagenumbering{alph}
\appendix

\section{Estimation method based on statistical measures}

\subsection{Difficulty of estimation by the fourth-order cumulant}
We argued that estimating the parameter $\Delta D$, which represents anisotropy, from a fourth-order cumulant is difficult. To clarify the quantitative reason for this, we consider the following simple model that assumes noiseless observations. Suppose that the position displacement $\Delta x_i$ follows a probability distribution 
\begin{align}
    \Delta x_i \sim \mathbf{Norm}(0,\sigma^2)
\end{align}
independently for each label $i$. Consider the distribution of sample variance and the sample fourth-order cumulant 
\begin{align}
    v &= \frac{1}{N}\sum_{i=1}^N  \Delta x_i^2 \label{v}\\
    d &=  \frac{1}{N} \sum_{i=1}^N  \Delta x_i^4 - 3\qty(\frac{1}{N}\sum_{i=1}^N  \Delta x_i^2)^2\label{d}
\end{align}
are computed from $N$ independent data.

Using mathematical induction, for even $n$, $\Delta x_i^n$ follows a generalized gamma distribution with $p=2/n, d=1/n, a = (2\sigma^2)^\frac{1}{n}$. This allows us to obtain 
\begin{align}
    \expval{\Delta x_i^2} &= \sigma^2 \\
    \expval{\Delta x_i^4 }&= 3\sigma^4 \\
    \expval{\Delta x_i^6 }&= 15\sigma^6 \\
    \expval{\Delta x_i^8} &= 105\sigma^8 .
\end{align}
Using these results for Eqs. [\ref{v}] and [\ref{d}], we obtain 
\begin{align}
    \expval{v} &= \sigma^2 \\
    \expval{(v-\expval{v})^2}&= \frac{2}{N} \sigma^4 \\
    \expval{d}&= -\frac{6}{N} \sigma^4 \\
    \expval{(d-\expval{d})^2} &= \qty(\frac{24}{N} -\frac{72}{N^2} +\frac{432}{N^3})\sigma^8.
\end{align}

Our model of anisotropic diffusion is more complex than this, but for simplicity, we assume that the position displacement follows a Gaussian distribution with $\sigma^2 = 2\bar{D} t$. This assumption is strictly valid in the short-time limit $t\to 0$ and long-time limit $t\to \infty$. Under this assumption, the sample variance is $8t^2\bar{D}^2/N$, and the variance of the sample cumulant is $384\bar{D}^4 t^4 / N + \mathcal{O}(1/N^2)$. These results do not agree with those in \cite{han2006brownian} but are consistent with those obtained by numerical simulation.

\section{Detailed description of EM algorithm}

In a model incorporating observational noises, 
%the update equations for EM algorithm are expressed as 
EM algorithm updates $D_a$, $D_b$, and $D_\theta$ are expressed as
\begin{align}
  D_a & \leftarrow \frac{1}{2T} \expval{\sum_{i=1}^N  (\cos{\theta_i}\Delta x_i + \sin{\theta_i} \Delta y_i)^2} \label{da} \\
  D_b & \leftarrow \frac{1}{2T} \expval{\sum_{i=1}^N (\sin{\theta_i}\Delta x_i -\cos{\theta_i}\Delta y_i)^2} \label{db} \\
  D_\theta &\leftarrow \frac{1}{2T} \expval{\sum_{i=1}^N \Delta \theta_i^2}  \label{dtheta}
\end{align}
where $T= N \Delta t$ and the expectation $\expval{\cdots}$ is taken with respect to
%under a parameter-fixed posterior distribution

\begin{align}
    p(\vb{x},\vb{y}, \boldsymbol{\theta}  \mid  \vb{X}, \vb{Y}, \vb{D}) \propto \prod_{i=1}^{N-1} \exp(-\frac{u_i^2}{4D_a\Delta t} -\frac{v_i^2}{4D_b\Delta t} - \frac{\Delta\theta_i^2}{4D_\theta \Delta t}) \prod_{i=1}^{N} \exp(-\frac{(X_i-x_i)^2}{2\epsilon^2} -\frac{(Y_i-y_i)^2}{2\epsilon^2}). \label{nonise} 
\end{align}

where $\vb{D}=(D_a,D_b,D_\theta)$, $u_i = \cos{\theta_i}\Delta x_i +\sin{\theta_i} \Delta y_i$, $v_i = \sin{\theta_i}\Delta x_i -\cos{\theta_i} \Delta y_i$ and $\Delta \theta_i = \theta_{i+1} - \theta_i$.
While $\vb{x} = \{x_i\}_{i=1}^N$, $\vb{y} = \{y_i\}_{i=1}^N$, and $\boldsymbol{\theta} =\{\theta_i\}_{i=1}^N$
represent variables for estimating a trajectory, $\vb{X} = \{X_i\}_{i=1}^N$ and $\vb{Y} = \{Y_i\}_{i=1}^N$ denote measurements of the single trajectory.

We handle the angular variable $\theta_i$ 
($i=1,\ldots,N$) as
being wrapped to $[0, 2\pi)$ for the convenience of estimation.
For each index $i$, the averages 
on the right-hand sides of Eqs. [\ref{da}], [\ref{db}], and [\ref{dtheta}] depend only on 
the marginalized joint posterior 
$p(\Delta x_i, \Delta y_i, \Delta \theta_i  \mid  \vb{X}, \vb{Y}, \vb{D})$, 
which can be efficiently assessed by belief propagation (BP)~\cite{pearl1988probabilistic}. 

For constructing the BP algorithm, we first decompose the joint distribution as
\begin{align}
    p(\Delta x_i, \Delta y_i, \Delta \theta_i  \mid  \vb{X}, \vb{Y}, \vb{D}) \propto
  \nu_i(\x_i)f(\x_i,\x_{i+1})\mu_{i+1}(\x_{i+1}) \label{joint1}
\end{align}
where $\x_i$ denotes the tuple of $x_i,y_i,\theta_i$. 
Conditonal distributions $\nu_i(\x_i)=p(\x_i \mid X_1, \dots, X_i, Y_1, \cdots, Y_i)$ and   $\mu_i(\x_i)=p(\x_i \mid X_i, \dots, X_N, Y_i, \cdots, Y_N)$ 
are termed forward and backward messages, respectively, and 
\begin{align}
    f(\x_i, \x_{i+1}) &\propto g(x_{i+1},y_{i+1},\x_i) h(\theta_{i+1},\theta_i) \\
    g(x_{i+1},y_{i+1},\x_i) &= \exp(-\frac{u_i^2}{4D_a\Delta t} -\frac{v_i^2}{4D_b\Delta t})  \\
    h(\theta_{i+1},\theta_i)  &= \exp(- \frac{\Delta\theta_i^2}{4D_\theta \Delta t}).
\end{align}
This means that one can assess $p(\Delta x_i, \Delta y_i, \Delta \theta_i  \mid  \vb{X}, \vb{Y}, \vb{D})$ by computing 
$\nu_i(\x_i)$ and $\mu_i(\x_i)$, which can be carried out in a recursive manner.

\subsection{Calculation of forward and backward messages}

\begin{figure}[!htb]
    \centering
    \includegraphics[width=0.5\linewidth]{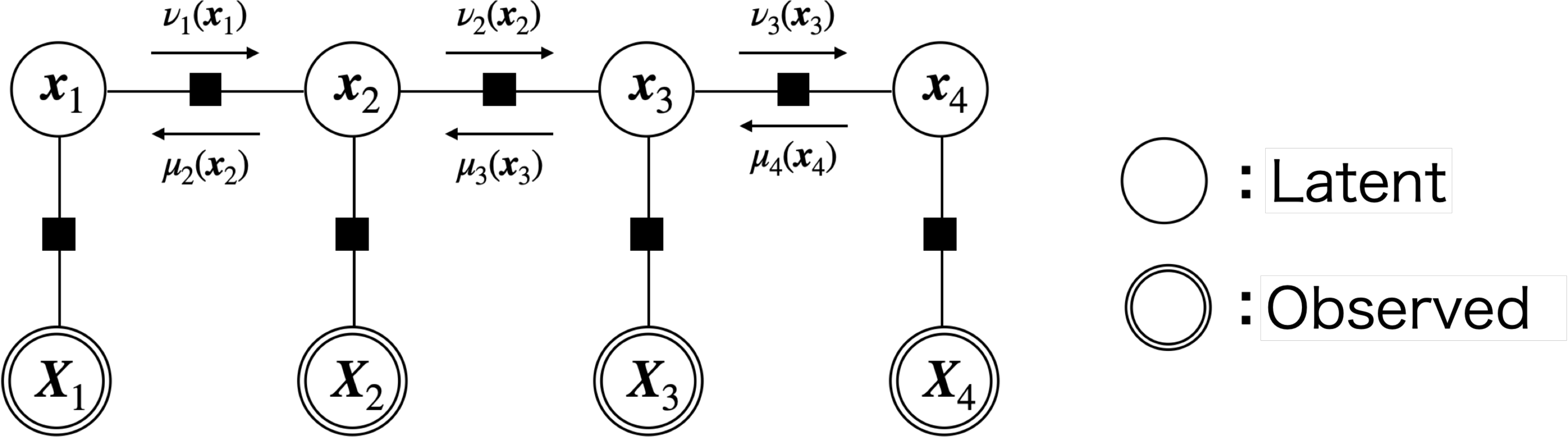}
    \caption{A factor graph of the posterior distribution in $N=4$.  The circles represent latent variables $\{ x_i,y_i,\theta_i \}$, while the double circles represent observed variables $\{ X_i,Y_i \}$. The square nodes represent the factor functions that relate the latent and observed variables in a factorized form. The $\nu_i$ and $\mu_i$ represent forward and backward messages respectively, which show the conditional probability distributions calculated from the known previous nodes. These messages are calculated recursively, and the information from these messages can be used to obtain the marginal probability distributions for each time step and between adjacent time steps.}  
    \label{fig:factor_graph}
\end{figure}
\FloatBarrier

More precisely, a recurrence formula

\begin{align}
    \nu_{i} (\x_{i}) &= \frac{1}{Z_i^{\rm F}} \int d\x_{i-1} ~ \nu_{i-1}(\x_{i-1}) p(\x_i \mid \x_{i-1})  p(x_i, y_i \mid X_i, Y_i)\\
    &= \frac{1}{Z_i^{\rm F}} \int d\x_{i-1} ~ \nu_{i-1}(\x_{i-1}) 
    \mathbf{Norm}(\theta_i \mid \theta_{i-1}, 2D_\theta \Delta t) 
    \mathbf{Norm}(x_i, y_i \mid \boldsymbol{\mu}_1, \vb{\Sigma}_1) 
    \mathbf{Norm}(x_i,y_i \mid \boldsymbol{\mu}_2, \boldsymbol{\Sigma}_2) 
    \label{eq:rec}
\end{align}

holds for $\nu_i(\x_i)$ (Figure \ref{fig:factor_graph}), 
where $Z_i^{\rm F}$ is a normalization constant, $\boldsymbol{\mu}_1=(X_i,Y_i)^\top$, 
$\boldsymbol{\mu}_2=(x_{i-1},y_{i-1})^\top$, $\vb{\Sigma}_1 = \epsilon^2 \vb{I}$,  and
\begin{align}
    \boldsymbol{\Sigma}_2=\mqty(2\bar{D}+\Delta D \cos{2\theta_i} & \Delta D\sin{2\theta_i} \\ \Delta D \sin{2\theta_i} & 2\bar{D}-\Delta D \cos{2\theta_i})\Delta t. 
\end{align}
A similar formula for computing $\mu_i(\x_i)$ from $\mu_{i+1}(\x_{i+1})$ also holds. 
%Normally, such a recurrence formula is calculated using propagation and resampling steps, which lead particles to degenerate in this model, because the space being explored is three-dimensional.  
Unfortunately, owing to the periodicity of the wrapped variable $\theta_i$, analytically assessing the right-hand side of Eq. [\ref{eq:rec}] is difficult. For practically resolving this difficulty, we resort to sampling methods known as "particle filters" or "sequential Monte Carlo methods"~\cite{kitagawa1996monte,liu1998sequential}. 

Particle filters express relevant distributions by populations of samples. The samples are termed "particles".  
There are several ways for implementing Eq. [\ref{eq:rec}] as a sampling algorithm. The one we chose was as follows. 
For $i=1$, we prepare many ($10^4 \sim$) particles by independent sampling 
from $\nu_1(\x_1) = \mathbf{Norm}(x_1,y_1 \mid
(X_1, Y_1)^\top, \epsilon^2 \vb{I} )$. 
For updating the population, we rewrite Eq. [\ref{eq:rec}] as

\begin{align}
    \nu_{i} (\x_{i}) = \frac{1}{Z_i^{\rm F}}\int d\x_{i-1}~ C(\x_{i-1})\nu_{i-1}(\x_{i-1})  \mathbf{Norm}(x_i,y_i \mid \boldsymbol{\mu}, \boldsymbol{\Sigma} ) \mathbf{Norm}(\theta_i \mid \theta_{i-1}, 2D_\theta \Delta t)\label{eq:rec_C}
\end{align}

where
\begin{align}
    C(\x_{i-1})&= \frac{1}{2\pi \sqrt{\abs
{\vb{\Sigma}_1+\vb{\Sigma}_2}}} \exp(-\frac{(\boldsymbol{\mu}_1-\boldsymbol{\mu}_2)^\top (\vb{\Sigma}_1+\vb{\Sigma}_2)^{-1} (\boldsymbol{\mu}_1-\boldsymbol{\mu}_2) }{2})\\
&= \frac{1}{2\pi \sqrt{\epsilon^4 + 2(D_a + D_b) \epsilon^2\Delta t + D_a D_b\Delta t^2}} \exp(-\frac{(\boldsymbol{\mu}_1-\boldsymbol{\mu}_2)^\top (\vb{\Sigma}_1+\vb{\Sigma}_2)^{-1} (\boldsymbol{\mu}_1-\boldsymbol{\mu}_2) }{2})\\
    \vb{\Sigma} &= \vb{\Sigma_1} (\vb{\Sigma}_1 + \vb{\Sigma}_2)^{-1} \vb{\Sigma}_2 \\
    \boldsymbol{\mu} &= \vb{\Sigma}_2 (\vb{\Sigma}_1 +\vb{\Sigma}_2)^{-1}\boldsymbol{\mu}_1 + \vb{\Sigma}_1 (\vb{\Sigma}_1+\vb{\Sigma}_2)^{-1} \boldsymbol{\mu}_2 .
\end{align}

The right-hand side of this expression indicates that for $i\ge 2$, 
one can evaluate the $i$th population by 
resampling $\x_{i-1}$ from the $i-1$th population according to the weight of $C(\x_{i-1})$, and sampling $\x_i$ from $\mathbf{Norm}(x_i,y_i \mid \boldsymbol{\mu}, \boldsymbol{\Sigma} ) \mathbf{Norm}(\theta_i \mid \theta_{i-1}, 2D_\theta \Delta t)$ after that. 
In this manner, the sample representation of $\nu_i(\x_i)$ is assessed for
$i=1,\ldots,N$. Similarly, that of $\mu_i(\x_i)$ is computed recursively 
from $i=N$ to $i=1$

\subsection{Calculation of the joint distribution: decoupling approximation}
Although Eq. [\ref{joint1}] looks simple at a glance, it is difficult to evaluate using particle filters. A straightforward approach would be
to sample a pair of $\x_i$ and $\x_{i+1}$ according to 
$\nu_i(\x_i)$ and $\mu_{i+1}(\x_{i+1})$, respectively, 
and then to resample them according to the weight of $f(\x_i, \x_{i+1})$. 
However, owing to the lack of measured values, $\theta_i$ and $\theta_{i+1}$ sampled from $\nu_i(\x_i)$ and $\mu_{i+1}(\x_{i+1})$ are almost uncorrelated. This yields large statistical fluctuations of $f(\x_i, \x_{i+1})$ making the accurate assessment of Eq. [\ref{joint1}] challenging.  

A key for overcoming this challenge is to construct a functional estimate of the backward message, $\tilde{\mu}_{i+1}(\x_{i+1})$, from particles (This difficulty can also be resolved by employing a functional estimation of the forward message $\nu_i(\x_i)$.). With $\tilde{\mu}_{i+1}(\x_{i+1})$, one can evaluate Eq. [\ref{joint1}] by sampling 
$\x_i$ and $\x_{i+1}$ from $\nu_i(\x_i)$ and $p(\x_{i+1}|\x_i)$, respectively, and giving them the weight of $\tilde{\mu}_{i+1}(\x_{i+1})$.  
Nevertheless, owing to the multidimensional nature of $\x_{i+1}$, it is practically difficult to accurately estimate $\tilde{\mu}_{i+1}(\x_{i+1})$ from $10^4\sim 10^5$ particles, which is the limit of what our computational resources can handle. 
For resolving this difficulty, we introduce a decoupling 
approximation
\begin{align}
    {\mu}_{i+1}(\x_{i+1})
    \simeq \tilde{\mu}_{i+1}(x_{i+1},y_{i+1}) \times \tilde{\mu}_{i+1}(\theta_{i+1}), 
\end{align}
where, based on particles, $\tilde{\mu}_{i+1}(\theta_{i+1})$ is assessed
by kernel density estimation and $\tilde{\mu}_{i+1}(x_{i+1},y_{i+1})$ is estimated as Gaussian. 
Under this approximation, Eq. [\ref{joint1}] is assessed as

\begin{align}
     p(\x_i, \x_{i+1}) &\propto \nu_i(\x_i) g(x_{i+1},y_{i+1},\x_i) h(\theta_{i+1},\theta_i) 
     \mu_{i+1}(\x_{i+1}) \label{eq:approx1} \\
     &\cong \frac{1}{M} \sum_{m=1}^M  g(x_{i+1},y_{i+1},\x_i^{(m)}) h(\theta_{i+1},\theta_i^{(m)}) {\mu}_{i+1}(\x_{i+1}) \notag\\
     & ~~~~~~ \times \delta(x_{i}-x_{i}^{(m)})\delta(y_{i}-y_{i}^{(m)}) \delta(\theta_i - \theta_i^{(m)}) \label{eq:approx2}\\
     &\cong \frac{1}{M} \sum_{m=1}^M  g(x_{i+1},y_{i+1},\x_i^{(m)}) h(\theta_{i+1},\theta_i^{(m)}) \tilde{\mu}_{i+1}(x_{i+1},y_{i+1})\tilde{\mu}_{i+1}(\theta_{i+1})\notag \\
     & ~~~~~~ \times \delta(x_{i}-x_{i}^{(m)})\delta(y_{i}-y_{i}^{(m)}) \delta(\theta_i - \theta_i^{(m)}))\label{eq:approx3} \\
     &\cong \frac{1}{M^2} \sum_{m=1}^M  \sum_{l=1}^M g(x_{i+1}^{(l)},y_{i+1}^{(l)},\x_i^{(m)}) h(\theta_{i+1},\theta_i^{(m)}) \tilde{\mu}_{i+1}(x_{i+1}^{(l)},y_{i+1}^{(l)})\tilde{\mu}_{i+1}(\theta_{i+1}) \notag\\
     & ~~~~~~ \times \delta(x_{i+1}-x_{i+1}^{(l)})\delta(y_{i+1}-y_{i+1}^{(l)}) \delta(x_{i}-x_{i}^{(m)})\delta(y_{i}-y_{i}^{(m)}) \delta(\theta_i - \theta_i^{(m)}) \label{eq:approx4}\\
     &\cong \frac{1}{M} \sum_{m=1}^M  g(x_{i+1}^{(m)},y_{i+1}^{(m)},\x_i^{(m)}) h(\theta_{i+1},\theta_i^{(m)}) \tilde{\mu}_{i+1}(x_{i+1},y_{i+1})\tilde{\mu}_{i+1}(\theta_{i+1}) \notag\\
     & ~~~~~~ \times \delta(x_{i+1}-x_{i+1}^{(m)})\delta(y_{i+1}-y_{i+1}^{(m)}) \delta(x_{i}-x_{i}^{(m)})\delta(y_{i}-y_{i}^{(m)}) \delta(\theta_i - \theta_i^{(m)}) \label{eq:approx5}\\
     &\cong \frac{1}{M} \sum_{m=1}^M   \tilde{\mu}_{i+1}(x_{i+1}^{(m)},y_{i+1}^{(m)})\tilde{\mu}_{i+1}(\theta_{i+1}^{(m)}) \notag\\
     & ~~~~~~ \times \delta(x_{i}-x_{i}^{(m)})\delta(y_{i}-y_{i}^{(m)}) \delta(\theta_i - \theta_i^{(m)})\delta(x_{i+1}-x_{i+1}^{(m)})\delta(y_{i+1}-y_{i+1}^{(m)}) \delta(\theta_{i+1} - \theta_{i+1}^{(m)}) \label{eq:approx6}
\end{align}
where 
$\vb{z}_i^{(m)} = (x_i^{(m)},y_i^{(m)}, \theta_i^{(m)})$ represents 
particles of the $i$th population of $\nu_i(\vb{z}_i)$ and 
$(x_{i+1}^{(m)}, y_{i+1}^{(m)}, \theta_{i+1}^{(m)})$ denotes 
particles sampled from $p(x_{i+1},y_{i+1},\theta_{i+1}\mid 
\x_i^{(m)})\propto g(x_{i+1},y_{i+1}, \x_i^{(m)})h(\theta_{i+1},\theta_i^{(m)})$. 
The decoupling approximation is employed from Eq. [\ref{eq:approx2}] to Eq. [\ref{eq:approx3}], 
and the Monte Carlo approximations are used from Eq. [\ref{eq:approx1}] to Eq. [\ref{eq:approx2}] 
and from Eq. [\ref{eq:approx3}] to Eq. [\ref{eq:approx6}].

\subsection{Calculation of the log-likelihood}
The EM algorithm locally maximizes the log-likelihood, and its value is used for selecting a suitable solution among multiple candidates. 
In the current system, the forward messages of BP can also be used for efficiently evaluating the log-likelihood. 
Let $L_n=\log{p(\vb{X}_1, \vb{X}_2, \cdots ,\vb{X}_n \mid \vb{D})}$ be the log-likelihood given $\vb{X}_1, \vb{X}_2, \cdots \vb{X}_n$. 
%The following recurrence relation holds from [\ref{eq:rec}]:
Bayes' theorem $p(\vb{X}_n\mid \vb{X}_1, \cdots, \vb{X}_{n-1}, \vb{D})
=p(\vb{X}_1, \vb{X}_2, \cdots, \vb{X}_n \mid \vb{D})
/p(\vb{X}_1, \vb{X}_2, \cdots, \vb{X}_{n-1} \mid \vb{D})$
offers a recurrence formula
\begin{align}
    L_n &=  L_{n-1} + \log{p(\vb{X}_n\mid \vb{X}_1, \cdots ,\vb{X}_{n-1}, \vb{D}})\\
        &=  L_{n-1} + \log\left(  \int d\x_n \int d\x_{n-1} p(\x_n\mid \x_{n-1}) \right.  p(\vb{X}_n\mid \x_n) p(\x_{n-1}\mid \vb{X}_1, \cdots \vb{X}_{n-1}, \vb{D}) \bigg) \\
    &\simeq L_{n-1} 
    %+ \log(\frac{1}{M}\sum_{m=1}^M C(x_{n-1}^{(m)},y_{n-1}^{(m)},\theta_{n-1}^{(m)})),
    + \log(\frac{1}{M}\sum_{m=1}^M C(\x_{n-1}^{(m)})),
\end{align}
where $M$ is the number of particles. 
This means that one can compute the log-likelihood as a byproduct of assessing the forward messages.

\section{Fixed point analysis of the EM algorithm}
\subsection{The noiseless and infinitely long trajectory limit regime: Analytical results}
In the main text, 
the same notation $\vb{D} = (D_a, D_b, D_\theta)$ is used for 
both the true diffusion coefficients and estimated variables depending on the context
to simplify notation.
However, for clarity of discussion, 
we denote  
the true diffusion parameters as $\DD=(\cD_a, \cD_b, \cD_\theta)$ 
to distinguish it from the estimated variables $\vb{D}$ in this section. 

When measurement noises are absent, $\vb{x}=\vb{X}$ and $\vb{y} = \vb{Y}$ hold. 
Then, the EM algorithm is expressed as
\begin{align}
  D_a & \leftarrow \frac{1}{2T} \expval{\sum_{i=1}^N  (\cos{\theta_i}\Delta X_i + \sin{\theta_i} \Delta Y_i)^2} \label{10} \\
  D_b & \leftarrow \frac{1}{2T} \expval{\sum_{i=1}^N (\sin{\theta_i}\Delta X_i -\cos{\theta_i}\Delta Y_i)^2} \label{11} \\
  D_\theta &\leftarrow \frac{1}{2T} \expval{\sum_{i=1}^N \Delta \theta_i^2}  
  \label{12}
\end{align}
where $\Delta X_i = X_{i+1}-X_i$, $\Delta Y_i = Y_{i+1} - Y_i$.
The expectation $\expval{\cdots}$ is assessed for 
% $p(\boldsymbol{\theta} \mid \vb{x}, \vb{y}, D_a, D_b, D_\theta)$, which is 
\begin{align}
    p(\boldsymbol{\theta} \mid \vb{X}, \vb{Y}, \vb{D}) \propto \prod_{i=1}^{N-1} \exp(-\frac{U_i^2}{4D_a\Delta t} -\frac{V_i^2}{4D_b\Delta t} - \frac{\Delta\theta_i^2}{4D_\theta \Delta t}), \label{nonoise} 
\end{align}
where we set $U_i = \cos \theta_i \Delta X_i + \sin \theta_i \Delta Y_i$ and 
$V_i = \sin \theta_i \Delta X_i - \cos \theta_i \Delta Y_i$.

Because of the invariance of Eq. [\ref{nonoise}] under the transformation of 
$(D_a, D_b, \{\theta_i\}) \to (D_b, D_a, \{\theta_i + \pi/2\})$, 
the EM algorithm always possesses 
an isotropic fixed point of $D_a = D_b = D^\ast$. 
Given the uniform distribution $p(\theta_1) = 1/(2\pi)$ for 
the angular variable at the initial time $i=1$, this solution guarantees that 
$\theta_i$ is also distributed uniformly over $[0,2\pi)$ 
for all of $i=2,\ldots,N$.
In the long trajectory limit of $N\to \infty$, this, in conjunction with Eqs. 
[\ref{10}] and [\ref{11}], yields
\begin{align}
    D_a = D_b=D^* = \frac{1}{4T} \sum_{i=1}^N  (\Delta X_i^2 + \Delta Y_i^2) \to  
    \bar{\cD} = \frac{\cD_a+\cD_b}{2}  ~~ (N\to \infty). \label{conv}
\end{align}

Next, to examine the stability of the fixed point, 
we expand Eqs. [\ref{10}], [\ref{11}], and [\ref{12}]
with respect to 
$\delta \vb{D}^{(n)} = (D_a^{(n)} -\bar{\cD}, 
D_b^{(n)}-\bar{\cD}, D_\theta^{(n)}-D_\theta^*)^\top$, 
which yields
\begin{align}
    \delta \vb{D}^{(n+1)} = L \delta \vb{D}^{(n)},
\end{align}
where $D_\theta^*$ is the value of $D_\theta$ at the fixed point and

\begin{eqnarray}
  L = \left (
  \begin{array}{ccc}
     \frac{1}{8T\bar{\cD}^2}
      (\ev{w_a^2} -\ev{w_a}^2) &  
      \frac{1}{8T\bar{\cD}^2}  
      (\ev{w_a w_b} - \ev{w_a}\ev{w_b} )& 
      \frac{1}{8T\bar{\cD}^2}  
      (\ev{w_a w_\theta} - \ev{w_a}\ev{w_\theta} )     
      \\
     \frac{1}{8T\bar{\cD}^2} 
      (  \ev{w_a w_b} -\ev{w_a}\ev{w_b}) &  
      \frac{1}{8T\bar{\cD}^2}  
      (\ev{w_b^2} - \ev{w_b}^2) & 
      \frac{1}{8T\bar{\cD}^2}  
      (\ev{w_b w_\theta} - \ev{w_b}\ev{w_\theta}) \\
    \frac{1}{8T(D_\theta^*)^2} 
      (  \ev{w_a w_\theta} -\ev{w_a}\ev{w_\theta}) &  
      \frac{1}{8T(D_\theta^*)^2}   
      (\ev{w_b w_\theta } - \ev{w_b}\ev{w_\theta}) & 
      \frac{1}{8T(D_\theta^*)^2}   
      (\ev{w_\theta^2} - \ev{w_\theta}^2)
  \end{array}
  \right ), \label{eq:L}
\end{eqnarray}

\begin{eqnarray}
\left (
    \begin{array}{c}
         w_a \\
         w_b \\
         w_\theta 
    \end{array}
  \right )
  = 
  \left (
  \begin{array}{c}
    \sum_i U_i^2\\
    \sum_i V_i^2\\
    \sum_i \Delta\theta_i^2
  \end{array}
\right ).
\end{eqnarray}
Eq.~[\ref{nonoise}] offers various moments necessary for assessing Eq. 
[\ref{eq:L}] as follows:

\begin{align}
  \ev{U^2_i} = \ev{V^2_i} = \frac{1}{2} \left(\Delta X_i^2+\Delta Y_i^2\right)
\end{align}

\begin{align}
  \ev{U^4_i} = \frac{1}{2\pi} \int_0^{2\pi} (\cos{\theta_i}\Delta X_i+\sin{\theta_i}\Delta Y_i)^4  d\theta_i  =\frac{3}{8}  \qty( \Delta X_i^2 + \Delta Y_i^2)^2
\end{align}

\begin{align}
  \ev{U^2_i U^2_j} &= \int_0^{2\pi}\int_{-\infty}^{\infty}\frac{(\cos{\theta_i}\Delta X_i+\sin{\theta_i}\Delta Y_i)^2(\cos{\theta_j}\Delta X_j+\sin{\theta_j}\Delta Y_j)^2\exp\left(-\frac{(\theta_i-\theta_j)^2}{4\abs{j-i} D_\theta^* \Delta t}\right)}{(2\pi)\sqrt{4\pi \abs{j-i} D_\theta^* \Delta t  } } d\theta_j d\theta_i \\
  &= \int_0^{2 \pi }\int_{-\infty }^{\infty }\frac{(\cos{\theta_i} \Delta X_i+\sin{\theta_i}\Delta Y_i)^2 (\cos (\theta_j+\theta_i) \Delta X_j+\sin (\theta_j+\theta_i) \Delta Y_j)^2 \exp \left(-\frac{\theta_j^2}{4 \abs{j-i} D_\theta^* \Delta t}\right)}{(2 \pi ) \sqrt{4 \pi  \abs{j-i} D_\theta^* \Delta t  }}d\theta_j d\theta_i\\
  &=\frac{1}{8} \bigg\{e^{-4 \abs{i-j} {D_\theta \Delta t}} \bigg(\left(\Delta X_i \left(\Delta X_j+\Delta Y_j\right)+\Delta Y_i \left(\Delta Y_j-\Delta X_j\right)\right) \notag \\ 
  &~~~~~~~~~~~~\left(\Delta X_i \left(\Delta X_j-\Delta Y_j\right)+\Delta Y_i \left(\Delta X_j+\Delta Y_j\right)\right)\bigg)+2 \left(\Delta X_i^2+\Delta Y_i^2\right) \left(\Delta X_j^2+\Delta Y_j^2\right) \bigg\} 
\end{align}

\begin{align}
  \ev{U^2_i V_i^2}  = \frac{1}{2\pi} \int_0^{2\pi} (\cos{\theta_i} \Delta X_i+\sin{\theta_i} \Delta Y_i)^2 (\sin{\theta_j} \Delta X_j-\cos{\theta_j} \Delta Y_j)^2 d\theta_i  =\frac{1}{8}  \qty( \Delta X_i^2 + \Delta Y_i^2)^2
\end{align}

\begin{align}
  \ev{U^2_i V^2_j} &=\int_0^{2\pi}\int_{-\infty}^{\infty}\frac{(\cos{\theta_i} \Delta X_i+\sin{\theta_i} \Delta Y_i)^2 (\sin{\theta_j} \Delta X_j-\cos{\theta_j} \Delta Y_j)^2 \exp\left(-\frac{(\theta_i-\theta_j)^2}{4 \abs{j-i} D_\theta^* \Delta t}\right)}{(2\pi)\sqrt{4\pi \abs{j-i} D_\theta^* \Delta t}}d\theta_j d\theta_i\\
  &=\int_0^{2\pi}\int_{-\infty}^{\infty}\frac{(\cos{\theta_i} \Delta X_i+\sin{\theta_i} \Delta Y_i)^2 (\sin(\theta_j+\theta_i) \Delta X_j-\cos(\theta_j+\theta_i) \Delta Y_j)^2 \exp\left(-\frac{\theta_j^2}{4\abs{j-i} D_\theta^* \Delta t}\right)}{(2\pi)\sqrt{4\pi \abs{j-i} D_\theta^* \Delta t}}d\theta_j d\theta_i\\
  &=\frac{1}{8} \bigg\{ -e^{-4\abs{i-j}D_\theta \Delta t}(\Delta X_i(\Delta X_j+\Delta Y_j)+\Delta Y_i(\Delta Y_j-\Delta X_j)) (\Delta X_i(\Delta X_j-\Delta Y_j)+\Delta Y_i(\Delta X_j+\Delta Y_j)) \notag\\
  &~~~~~~~~~~~~+2(\Delta X_i^2+\Delta Y_i^2)(\Delta X_j^2+\Delta Y_j^2) \bigg\}
\end{align}

\begin{align}
  \ev{V^4_i} = \frac{1}{2\pi} \int_0^{2\pi} (\sin{\theta_i}\Delta X_i-\cos{\theta_i}\Delta Y_i)^4  d\theta_i  =\frac{3}{8}  \qty( \Delta X_i^2 + \Delta Y_i^2)^2
\end{align}
  
\begin{align}
  \ev{U^2_i V^2_j} &= \int _0^{2 \pi }\int _{-\infty }^{\infty }\frac{(\sin \theta_i \Delta X_i - \cos \theta_i \Delta Y_i)^2 (\sin \theta_j \Delta X_j - \cos \theta_j \Delta Y_j)^2 \exp \left(-\frac{(\theta_i - \theta_j)^2}{4 \abs{j-i} D_\theta^* \Delta t}\right)}{(2 \pi) \sqrt{4 \pi \abs{j-i} D_\theta^* \Delta t}}d\theta_j d\theta_i \\
  &=\int _0^{2 \pi }\int _{-\infty }^{\infty }\frac{(\sin \theta_i \Delta X_i - \cos \theta_i \Delta Y_i)^2 (\sin (\theta_j + \theta_i) \Delta X_j - \cos (\theta_j + \theta_i) \Delta Y_j)^2 \exp \left(-\frac{\theta_j^2}{4 \abs{j-i} D_\theta^* \Delta t }\right)}{(2 \pi) \sqrt{4 \pi \abs{j-i} D_\theta^* \Delta t}}d\theta_j d\theta_i\\
  &=\frac{1}{8} \bigg\{e^{-4 \abs{j-i} D_\theta \Delta t} (\Delta X_i (\Delta X_j + \Delta Y_j) + \Delta Y_i (\Delta Y_j - \Delta X_j)) (\Delta X_i (\Delta X_j - \Delta Y_j) + \Delta Y_i (\Delta X_j + \Delta Y_j)) \notag\\
  &~~~~~~~~~~~ + 2 (\Delta X_i^2 + \Delta Y_i^2) (\Delta X_j^2 + \Delta Y_j^2) \bigg\}
\end{align}

\begin{align}
    \ev{U^2_i \Delta \theta^2_j} =\ev{V^2_i \Delta \theta^2_j} =0
\end{align}

\begin{align}
  \ev{\Delta \theta^2_i \Delta \theta^2_i} - \ev{\Delta \theta^2_i}\ev{ \Delta \theta^2_i} &= 0\\
  \ev{\Delta \theta^2_i \Delta \theta^2_j} - \ev{\Delta \theta^2_i}\ev{ \Delta \theta^2_j} &=  8 D_\theta^{2} \Delta t^2 .
\end{align}
Based on these, $L$ for 
a given trajectory $\vb{X}, \vb{Y}$
is computed as 
\begin{align}
  L=\frac{1}{2} \mqty(q & -q & 0 \\ -q & q & 0 \\ 0 & 0 & 2)
\end{align}
where 
\begin{align}
  q &= \frac{1}{32N\bar{\cD}^{2} \Delta t^2}  \sum_{ij}  \bigg\{e^{-4 \abs{i-j} {D_\theta^\ast \Delta t}} \bigg(\left(\Delta X_i \left(\Delta X_j+\Delta Y_j\right)+\Delta Y_i \left(\Delta Y_j-\Delta X_j\right)\right) \left(\Delta X_i \left(\Delta X_j-\Delta Y_j\right)+\Delta Y_i \left(\Delta X_j+\Delta Y_j\right)\right)\bigg) \bigg\}   . \label{same} 
\end{align}

Stability matrix $L$ has three eigenvalues $0, 1$, and $q$. 
The first two, $0$ and $1$, make the fixed 
point stable and marginally stable in the directions of $(1,1,0)^\top$ and $(0,0,1)^\top$, respectively. 
However, the last one, $q$, which corresponds to 
the eigenvector of $(1,-1,0)^\top$, destabilizes the fixed point  
breaking the symmetry of $D_a = D_b$ when $q > 1$. 
In the limit of $N\to \infty$, the law of large numbers guarantees that 
$q$ converges to its average with respect to the true distribution $p(\vb{X}, \vb{Y}\mid \DD)$. 
The moments of $\{\Delta X_i, \Delta Y_i\}_{i=1}^N$ necessary for computing 
the average of $q$ are assessed as 

\begin{align}
  [\Delta X_i^2\Delta X_j^2]&=(\cD_a+\cD_b)^2 \Delta t^2 +\frac{1}{2} \Delta \cD^2 \Delta t^2 \exp(-4\cD_\theta \Delta t\abs{i-j} )\\
  [\Delta X_i^4]&=\frac{3}{2} (3\cD_a^2 + 2\cD_a \cD_b +3\cD_b^2) \Delta t^2\\
  [\Delta X_i^2\Delta Y_j^2]&=(\cD_a+D_b)^2\Delta t^2-\frac{1}{2} \Delta \cD^2 \Delta t^2 \exp(-4\cD_\theta \Delta t\abs{i-j} )\\
  [\Delta X_i^2\Delta Y_i^2]&= \frac{3}{2} (\cD_a^2+\cD_b^2)\Delta t^2 +\cD_a \cD_b \Delta t^2\\
  [\Delta X_i\Delta X_j \Delta Y_i \Delta Y_j]&=\frac{1}{2}\Delta \cD^2 \Delta t^2  \exp(-4\cD_\theta \Delta t\abs{i-j} )
\end{align}
where $[\cdots]$ denotes the average with respect to $p(\vb{X}, \vb{Y}\mid \DD)$. 
These allow us to express the average of $q$ as

\begin{align}
  [q] &= \frac{1}{32N\bar{\cD}^2 }\qty{\sum_i^N (12 \cD_a^2 + 8\cD_a \cD_b + 12\cD_b^2) + 4\Delta \cD^2\sum_{i\neq j}^N   \exp(-4\cD_\theta \Delta t \abs{i-j})  }  \\
    &= \frac{1}{8\bar{\cD}^2}\qty(3 \cD_a^2 + 2\cD_a \cD_b + 3\cD_b^2 + 2\Delta \cD^2 \frac{1}{e^{4\cD_\theta \Delta t}-1}) +  \mathcal{O}(N^{-1})\\
    &= 1+  \qty(\frac{\cD_a-\cD_b}{\cD_a+\cD_b})^2 \qty(\frac{1}{e^{4\cD_\theta \Delta t}-1} +\frac{1}{2})+  \mathcal{O}(N^{-1}),  \label{q}
\end{align}
which is always greater than $1$.
This means that as long as $\cD_a\neq \cD_b$ holds for the true parameters, 
the isotropic fixed point, which indicates the failure of anisotropy detection, 
becomes unstable when measurement noises are absent and the length of the trajectory tends to infinity. In other words, anisotropy can always be detected in this limit. 

\subsection{Validation of theoretical predictions using Monte Carlo simulations}
\begin{figure}[!htb]
    \centering
    \includegraphics[width=0.5\linewidth]{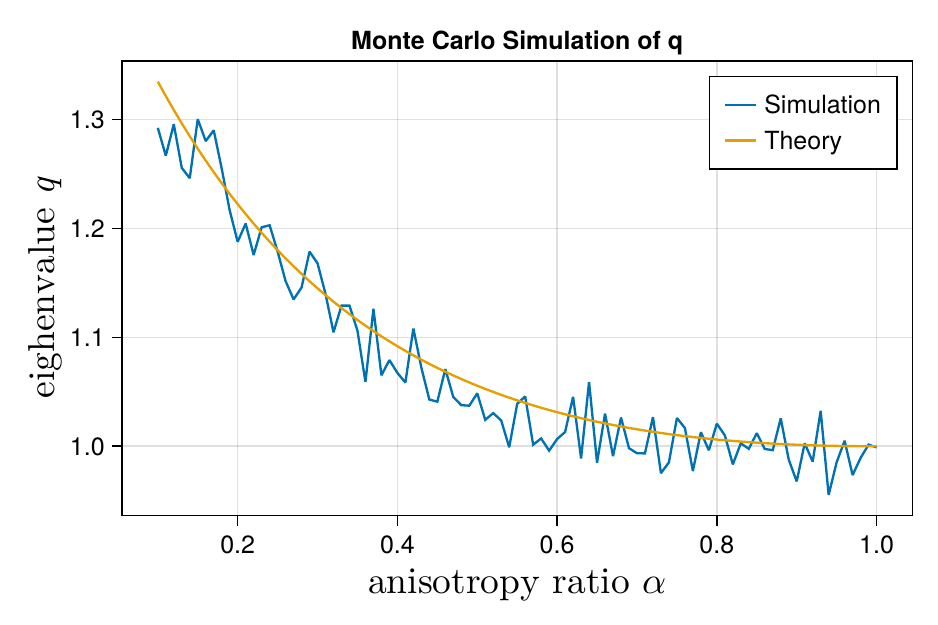}
    \caption{Sample-averaged eigenvalue $q$ of evolution matrix $L$ compared between the Monte Carlo simulation and analytical result with $D_a + D_b = 10.0$.}
    \label{fig:q_sim}
\end{figure}

To confirm the results of Eq. [\ref{q}], we computed the eigenvalue of $L$ directly from Eq. [\ref{eq:L}] using Monte Carlo methods and compared them with the results of Eq. [\ref{q}]. Figure \ref{fig:q_sim} shows 
how the eigenvalue $q$ depends on the anisotropy ratio $\alpha=\cD_b/\cD_a$ when $\cD_a+\cD_b=10$ and $\cD_\theta \to \infty$. This result indicates that the eigenvalues evaluated using the Monte Carlo methods are consistent with the theoretical assessment using Eq. [\ref{q}] giving eigenvalues greater than unity.

\subsection{Noisy and finite length regime: Numerical methods}
The analysis remains entirely the same up to Eq. [\ref{same}]. 
For simplicity, let us focus on the case of $D_\theta \to \infty$. 
In this case, the critical eigenvalue for a given measurement $\vb{X}, \vb{Y}$ is expressed as 
\begin{align}
    q = \frac{1}{32N{D^\ast}^2 \Delta t^2} \sum_i \expval{(\Delta x_i^2 + \Delta y_i^2)^2} 
    = \frac{1}{32N{D^\ast}^2 \Delta t^2} \sum_i \qty(\expval{\Delta x_i^4} + 2 \expval{\Delta x_i^2} \expval{\Delta y_i^2}+ \expval{\Delta y_i^4}) .  \label{q2}
\end{align}
where $D^*$ is the value of $D_a$ and $D_b$ for the isotropic fixed point and $\expval{\cdots}$ denotes the average with respect to $p(\vb{x},\vb{y} \mid \vb{X}, \vb{Y}, \vb{D^\ast})$. 
Note that $D^* = \bar{\cD}$ does not hold, and we need to numerically assess $D^*$ for each sample of $\vb{X}$ and $\vb{Y}$ using the EM algorithm
when the length of the trajectory is finite. 
The second equality of Eq. [\ref{q2}] holds as $\vb{x}$ and $\vb{y}$ become independent thanks to the condition $D_a=D_b = D^\ast$. 
Consequently, the assessment of $q$ is reduced to that for a one-dimensional variable model with respect to $\vb{x}$. In other words, our primary objective is to compute $p(\vb{x} \mid \vb{X}, \vb{D^\ast})$. This is carried out by the use of BP, where the forward and backward messages are computed recursively as  
\begin{align}
    \nu_i(x_i) &= \frac{1}{Z_i^{\rm F}} \int dx_{i-1} \nu_{i-1}(x_{i-1}) \mathbf{Norm}(x_i\mid x_{i-1}) \mathbf{Norm}(x_i\mid X_i), \label{rec_1df}\\
    \nu_i(x_i) &=  \mathbf{Norm}(x_1\mid X_1),\\
    \mu_i(x_i) &= \frac{1}{Z_i^{\rm B}} \int dx_{i+1} \mu_{i+1}(x_{i+1}) \mathbf{Norm}(x_i\mid x_{i+1}) \mathbf{Norm}(x_i\mid X_i),    \label{rec_1db} \\
    \mu_i(x_N) &=  \mathbf{Norm}(x_N\mid X_N).
\end{align}
After $\nu_i(x_i)$ and $\mu_{i+1}(x_{i+1})$ are provided, the joint distribution of two 
adjacent latent variables is computed as 
\begin{align}
    p(x_i,x_{i+1} ) \propto \nu_i(x_i) \mathbf{Norm}(x_{i+1}\mid x_i) \mu_{i+1} (x_{i+1}) .\label{rec_1dj}
\end{align}

Fortunately, as all the distributions under consideration are Gaussian, one can analytically perform BP by analytically
propagating its mean and variance as messages. Let us parameterize each message as 
\begin{align}
    \nu_i(x_i) &= \mathbf{Norm}(x_i\mid \mu_i^{\rm F}, (\sigma_i^{\rm F})^2), \label{f_n}\\
    \mu_i(x_i) &= \mathbf{Norm}(x_i\mid \mu_i^{\rm B}, (\sigma_i^{\rm B})^2), \label{b_n}\\
    p(x_i, x_{i+1}) &=  \mathbf{Norm}(x_i, x_{i+1} \mid \boldsymbol{\mu}_i, \vb{\Sigma}_i).  \label{two_joint}
\end{align}
Substituting Eqs. [\ref{f_n}], [\ref{b_n}], [\ref{two_joint}] into Eqs. [\ref{rec_1df}], [\ref{rec_1db}], and [\ref{rec_1dj}]  yields
\begin{align}
    ({\sigma_{1}^{\rm F}})^2 &= \epsilon^2 \\
    ({\sigma_{i+1}^{\rm F}})^2 &= \frac{\epsilon^2(\sigma^2+{(\sigma_{i}^{\rm F}})^2)}{\sigma^2+\epsilon^2+{\epsilon_i^{\rm F}}^2}  \\
    \mu^{\rm F}_{1} &= X_1 \\
    \mu^{\rm F}_{i+1} &= \frac{(\sigma^2 + {(\sigma_{i}^{\rm F}})^2)X_{i+1} + \epsilon^2 \mu_i^{\rm F}}{\sigma^2 + \epsilon^2 + ({\sigma_{i}^{\rm F}})^2 } \\
    ({\sigma_{N}^{\rm B}})^2 &= \epsilon^2 \\
    ({\sigma_{i}^{\rm B}})^2 &= \frac{\epsilon^2(\sigma^2+({\sigma_{i+1}^{\rm B}})^2)}{\sigma^2+\epsilon^2+({\sigma_{i+1}^{\rm B}})^2}  \\
    \mu^{\rm B}_{N} &= X_N \\
    \mu^{\rm B}_{i} &= \frac{(\sigma^2 + ({\sigma_{i+1}^{\rm B}})^2)X_{i} + \epsilon^2 \mu_{i+1}^{\rm B}}{\sigma^2 + \epsilon^2 + ({\sigma_{i+1}^{\rm B}})^2 }
\end{align}
where $\sigma^2 = 2D^\ast \Delta t$, and
\begin{align}
    \vb{\Sigma}_i &= \frac{1}{({\sigma_{i}^{\rm F}})^2 + ({\sigma_{i+1}^{\rm B}})^2+\sigma^2} 
    \mqty(({\sigma_{i}^{\rm F}})^2 ({\sigma_{i+1}^{\rm B}})^2 + \sigma^2({\sigma_{i}^{\rm F}})^2 
    && ({\sigma_{i}^{\rm F}})^2({\sigma_{i+1}^{\rm B}})^2 \\ 
    ({\sigma_{i}^{\rm F}})^2 ({\sigma_{i+1}^{\rm B}})^2 && 
    ({\sigma_{i}^{\rm F}})^2({\sigma_{i+1}^{\rm B}})^2 + \sigma^2({\sigma_{i+1}^{\rm B}})^2) 
    \equiv \mqty(\Sigma_{11}  \Sigma_{12} \\ \Sigma_{21}  \Sigma_{22})\\
    \boldsymbol{\mu}_i &= \frac{1}{({\sigma_{i}^{\rm F}})^2 + ({\sigma_{i+1}^{\rm B}})^2+\sigma^2} \mqty((({\sigma_{i+1}^{\rm B}})^2+\sigma^2)\mu_i^{\rm F} + ({\sigma_{i}^{\rm F}})^2\mu_{i+1}^{\rm B} \\ (({\sigma_{i}^{\rm F}})^2+\sigma^2)\mu_{i+1}^{\rm B} + ({\sigma_{i+1}^{\rm B}})^2\mu_{i}^{\rm F}) 
    \equiv \mqty(\mu_i \\ \mu_{i+1}). 
\end{align}
Using the notation above, one can evaluate the moments necessary for computing $q$ as follows:
\begin{align}
    \expval{\Delta x_i^2} &= \expval{(x_{i+1}-x_i)^2} \\
    &= \expval{\qty{(x_{i+1}-\mu_{i+1}) - (x_i - \mu_i) + (\mu_{i+1}-\mu_i)}^2} \\
    &= \expval{(x_{i+1}-\mu_{i+1})^2} + \expval{(x_{i}-\mu_{i})^2} - 2\expval{(x_{i+1}-\mu_{i+1})(x_i-\mu_i)} + (\mu_{i+1}-\mu_i)^2 \\
    &= \Sigma_{11} +\Sigma_{22} -2\Sigma_{12} + (\mu_{i+1}-\mu_i)^2 \\
    \expval{\Delta x_i^4} &= \expval{(x_{i+1}-x_i)^4} \\
    &= \expval{\qty{(x_{i+1}-\mu_{i+1}) - (x_i - \mu_i) + (\mu_{i+1}-\mu_i)}^4} \\
    &= \expval{(x_{i+1}-\mu_{i+1})^4} - 4\expval{(x_{i+1}-\mu_{i+1})^3(x_i-\mu_i)} 
      + 6\expval{(x_{i+1}-\mu_{i+1})^2(x_i-\mu_i)^2} \notag \\
    &~~~~~  + 6 (\mu_{i+1}-\mu_i)^2\expval{(x_{i+1}-\mu_{i+1}]^2}  -4\expval{(x_{i+1}-\mu_{i+1})(x_i-\mu_i)^3} \notag \\
    &~~~~~  -12 (\mu_{i+1}-\mu_i)^2\expval{(x_{i+1}-\mu_{i+1})(x_{i}-\mu_{i})} + \expval{(x_{i}-\mu_{i})^4} + 6(\mu_{i+1}-\mu_i)^2\expval{(x_{i}-\mu_{i})^2} + (\mu_{i+1}-\mu_i)^4 \\
    &= 3(\Sigma_{11}^2 +\Sigma_{22}^2) -12\Sigma_{12}(\Sigma_{11}+\Sigma_{22}) + 6(\mu_{i+1}-\mu_i)^2(\Sigma_{11}+\Sigma_{22}-2\Sigma_{12})\notag\\
    &~~~~~ + 6(\Sigma_{11}\Sigma_{22} + 2\Sigma_{12}^2)+(\mu_{i+1}-\mu_i)^4 .
\end{align}
After computing these, one can numerically evaluate the eigenvalue $q$ for a given measurement $\vb{X}, \vb{Y}$ by substituting it into Eq. [\ref{q2}].

\section{Characterization of the failure solution}
\begin{figure}[!htb]
    \centering
    \includegraphics[width=1.0\linewidth]{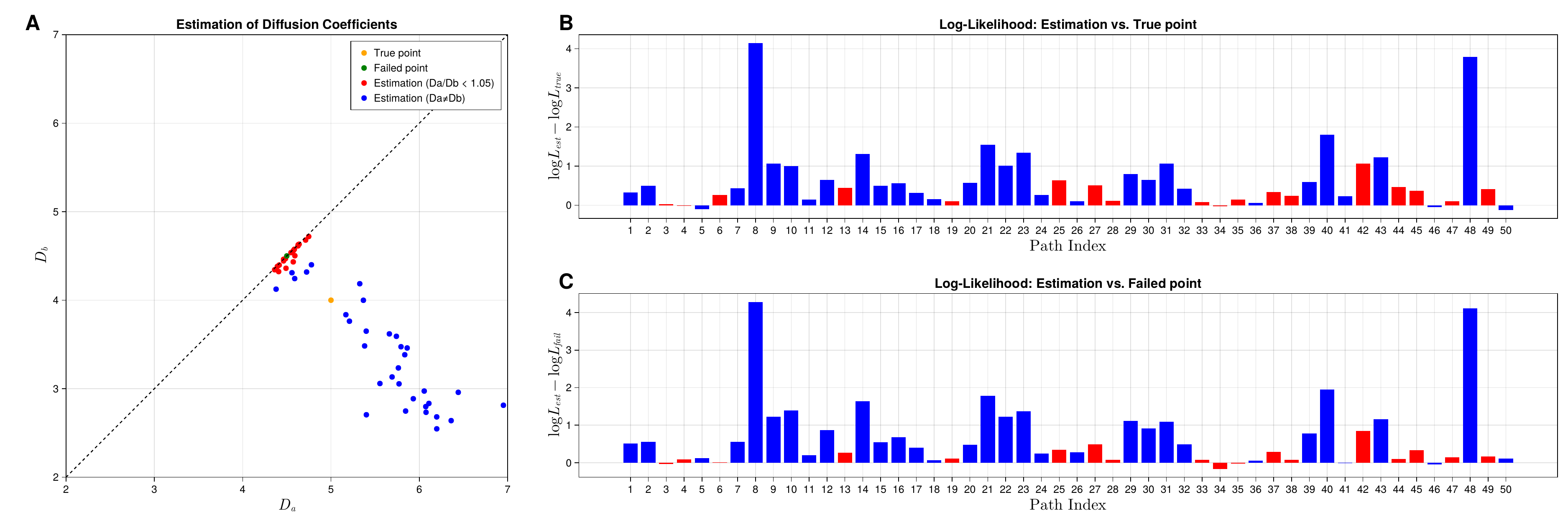}
    \caption{Comparison of log-likelihood for three candidates with local minima: (a) solution using the EM algorithm, (b) $\DD$ (true diffusion parameters), and (c) isotropic failure solution ($D_a=D_b=\bar{D}$). (A) Diffusion coefficient estimates obtained using the EM algorithm. (B) Comparison of log-likelihoods between (a) and (b).  (C) Comparison of log-likelihoods between (a) and (c). (B, C) Blue bars represent paths for which anisotropy is detected correctly, whereas red bars indicate paths for which the detection of anisotropy is failed. (A-C) $D_a=5.0, D_b=4.0, D_\theta = 10^5, \Delta t = 0.01, \epsilon = 0.1$. }
    \label{fig:loglike}
\end{figure}

Although the EM algorithm is guaranteed to converge to a local maximum, whether the obtained solution corresponds to the global maximum remains unclear. In other words, the true maximum likelihood estimate may lie elsewhere
even when the EM algorithm converges to the isotropic "failure solution". To distinguish between the two cases, we 
carried out computer experiments and compared the values of likelihood among three solutions: (a) the solution obtained using the EM algorithm, (b) the true diffusion parameters $\DD$, and (c) the isotoropic failure solution where $D_a=D_b=\bar{D}$ holds. Figure \ref{fig:loglike}B shows the difference of likelihood between (a) and (b) for 50 paths.
%for which anisotropy is correctly detected. 
This indicates that except for a few cases, the likelihood of (a) is greater than that of (b). 
Further, Figure \ref{fig:loglike}C shows the difference between (a) and (c) for 50 paths.
%for which the detection of anisotropy failed. 
In this case as well, the likelihood of (a) is greater than that of (c) except for a few cases. These results suggest that the local optima obtained using the EM algorithm correspond to the global optima in the current system with high probabilities.

\FloatBarrier

\section{Sample preparation}
\subsection{Fluorescent spheres}
The fluorescent spheres solution (F-8764, Invitrogen; 2$\%$ solids, nominal diameter 200 nm) was diluted $10^4$ times with Milli-Q water.  
10 $\mu$l of the diluted solution was mounted on a cover glass (No.1 22$\times$22 mm, Matsunami) and sealed by placing another cover glass on it with a double-sided PET tape of 5 $\mu$m-thickness (No.5600, Nitto) as a spacer between the two cover glasses.

\subsection{Quantum rods}
The CdSe/CdS core-shell type quantum rods (900514-1ML, Sigma-Aldrich) were diluted in toluene (204-17915, Wako) at a density of 0.5 $\mu$g/ml and
sonicated for 30 minutes (M1800-J, Branson) to disperse uniformly in suspension.
10 $\mu$l of the diluted sample was mounted on a cover glass (No.1 22$\times$22 mm, Matsunami) and sealed by placing another cover glass on it with a double-sided polyimide tape of 130 $\mu$m-thickness (Kincsem110-02, HCP) as a spacer.
The nominal major and minor axis length of the quantum rod was 28.4$\pm$3.0 and 4.6$\pm$0.7 nm, respectively.

\subsection{Bacteria}
$\it E. coli$ strain RA1 suspended in buffer (50 mM HEPES, pH 7.0) was first sterilized by heating at $60^\circ$C for 45 minutes  to terminate its swimming motion and observe only its pure diffusion. Then it was mounted on a cover glass (No.1 24$\times$60 mm, Matsunami). To observe quasi-two-dimensional diffusion of the bacteria, the resulting sample was sealed by placing another cover glass on it, with a double-sided PET tape of 5 $\mu$m-thickness (No.5600, Nitto) as the spacer.

\section{Microscopy}

\subsection{Fluorescent spheres}
Imaging of fluorescent microspheres was performed using an inverted microscope (IX83, Olympus) equipped with a 100 $\times$ oil-immersion objective (UPlanSAPO, Olympus), a mercury light source system (U-HGLGPS, Olympus) and a fluorescence filter unit (U-FBNA, Olympus).
The images were detected using a sCMOS camera (ORCA Flash4.0, Hamamatsu) at 100 fps with 1152 $\times$ 1152 pixels in 16-bit depth. The image acquisition process was controlled using cellSens (Olympus).

\subsection{Quantum rods}
The fluorescent images of quantum rods were captured using an inverted microscope (IX83, Olympus) equipped with a 100 $\times$ oil-immersion objective (UPlanSAPO, Olympus), a mercury light source system (U-HGLGPS, Olympus) and a fluorescence filter unit (U-FGNA, Olympus).
The images were detected using a sCMOS camera (ORCA Flash4.0, Hamamatsu) at 100 fps with 1152 $\times$ 1152 pixels in 16-bit depth. The image acquisition process was controlled using cellSens (Olympus).  To track the quantum rods moving in quasi-two dimensions whereever possible, fluorescent spots that were in focus during observation were selected to analyze their trajectories.

\subsection{Bacteria}
The transmitted images of bacteria were acquired using an inverted microscope (IX73, Olympus) equipped with a 100 $\times$ oil-immersion objective (UAPON100XOTIRF, Olympus) and transmitted illumination (M660L4, Thorlab). The images were detected using a high-speed camera (FASTCAM NovaS, Photron) at 3000 fps with 1024 $\times$ 1024 pixels in 8-bit depth.

\section{Image analysis}
 For each pixel, a temporal median was first calculated to obtain a background image which only contained tracers immobilized on the cover glass. The background image was then subtracted from each frame of the original image sequences, resulting in an image that only contained mobile tracers. The centroid of each fluorescent sphere and quantum rod at every time point was determined by fitting an isotropic Gaussian profile onto the respective spot image.
The centroid and angle of each bacterium were calculated using an anisotropic Gaussian profile instead.
The measurement error $\epsilon$ was estimated as the positional standard deviation of the tracers immobilized on the glass wall.

\FloatBarrier

\end{document}